\documentclass[vecphys]{svmult}

\usepackage{makeidx}         
\usepackage{graphicx}        
\usepackage{multicol}        

\makeindex             

\newcommand{\up}{\uparrow}
\newcommand{\dw}{\downarrow}
\newcommand{\bra}{\langle}
\newcommand{\ket}{\rangle}

\newcommand{\pa}{\partial}
\newcommand{\Vect}{\mathbf}
\newcommand{\ve}{\varepsilon}

\begin{document}

\title*{Nonequilibrium Quantum Breakdown in a Strongly Correlated Electron System}
\author{Takashi Oka, Hideo Aoki}
\institute{
Department of Physics, University of Tokyo, Hongo, Tokyo 113-0033, 
Japan
\texttt{oka@cms.phys.s.u-tokyo.ac.jp, aoki@cms.phys.s.u-tokyo.ac.jp}
\footnote[0]{to be published in ``Quantum and Semi-classical Percolation \& Breakdown",\\
Lecture Notes in Physics (LNP),  Springer-Verlag}}
%
%
\maketitle

\section{Introduction}
During the past decades, there has been an 
increasing fascination and surprises with 
diverse {\it quantum many-body effects}.
With the magical touch of interaction 
a simple electron system may assume 
insulating, metallic, magnetic or 
superconducting states according as the 
control parameters are changed.  
Strongly correlated electron systems, as 
exemplified by the high-Tc superconductors 
and their host materials realized in 
transition-metal oxides, as well as by 
organic metals, have provided us with an ideal playground,
where various crystal structures with 
band-filling control and band-width control etc 
provide the richness in the phase diagram\cite{Imada1998}.

On the other hand, there is a long history of the interests 
in {\it non-equilibrium phase transitions}.  
Statistical mechanically, there is an intriguing 
problem of how we can generally 
define the notion of a ``phase" in non-equilibrium 
systems, but we can still discuss individual systems in specified 
non-equilibrium conditions to extract more general viewpoints.  
Now, if we combine the above two ingredients, namely, 
if we consider {\it strongly-correlated electron systems in 
non-equilibrium}, we plunge into an even more fascinating 
physics.  In fact, recent years have witnessed 
an upsurge of interests in non-equilibrium states in 
many-body systems with drastic changes in the 
electronic states in strong dc electric fields, 
in intense laser fields, etc.

\begin{figure}[thb]
\centering 
\includegraphics[width=6.cm]{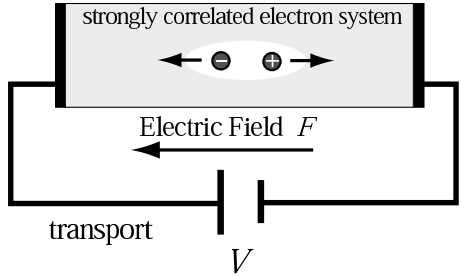}
\caption{
Non-linear transport and optical response
}
\label{title}
\end{figure}
Developments in fabrication techniques such as 
realization of clean thin films with electrodes attached 
have triggered several groundbreaking experiments, 
e.g., non-linear transport measurements in 
thin films \cite{Asamitsu1997,Ponnambalam1999,Oshima1999,
Liu2000, Baikalov2003,Sawa2004}, in layered systems \cite{Inagaki2004}
and observations of clean metallic states in 
heterostructures \cite{OhtomoNature}. 
Non-linear phenomena in correlated electron systems
now begin to attract interests in a wide range of 
researchers: One obvious area of application is future-generation 
electronic devices, where a high sensitivity of a system 
near a phase boundary to external conditions 
may lead to drastic functionalities\cite{Asamitsu1997}. 
However, even more attractive is its relevance to
fundamental physics, especially, to non-equilibrium statistical 
physics, where we can 
observe the behavior of various phase transitions 
taking place under non-equilibrium conditions.

The purpose of the present article is 
to discuss the nonequilibrium 
metal-insulator transition in 
strongly correlated electron systems \cite{Oka2005a,Oka2003,Oka2004a,Oka2004b}, 
which is known, for equilibrium systems, as Mott's transition.
Before going into detail, we first give a brief introduction
of the transition, and discuss how 
quantum breakdown through non-adiabatic transitions in 
nonequilibrium becomes relevant in non-linear transports.

The model we study
is the single-band Hubbard model, which is the simplest 
possible one that captures many essential properties of 
correlated electron physics.  The Hamiltonian reads
\begin{eqnarray}
H_0=-t_{{\rm hop}}\sum_{\bra i,j\ket \sigma}\left(c^\dagger_{i\sigma}c_{j\sigma}+\mbox{h.c.}\right)
+U\sum_{i}n_{i\up}n_{i\dw},
\end{eqnarray}
where $c_{i\sigma}$ annihilates an electron 
on site $i$ with spin $\sigma$, $n_{i\sigma}=
c^\dagger_{i\sigma}c_{i\sigma}$ the number operator, $U$ 
the strength of the on-site Coulomb repulsion, 
and $t_{{\rm hop}}$ the hopping integral. 
The filling $\displaystyle{n=\frac{1}{L}\sum_{i=1}^L\bra 
n_{i\up}+n_{i\dw} \ket}$,
with $L$ the number of sites, is an important parameter, which 
changes the groundstate property drastically.

\begin{figure}[thb]
\centering 
\includegraphics[width=10.cm]{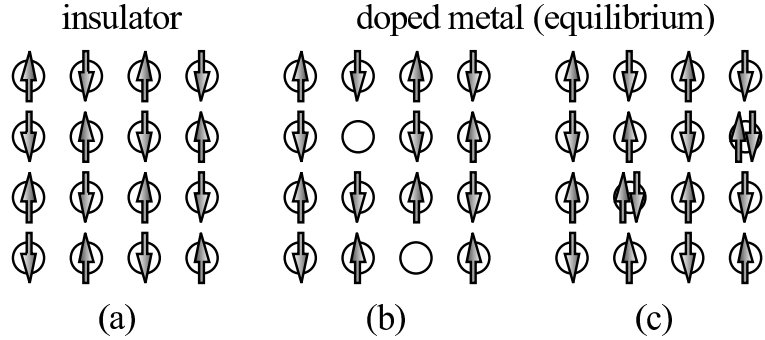}
\caption{Metal-insulator transition in equilibrium due to doping:
(a) A Mott insulator realized at half filling.
(b) A hole-doped metallic state. 
(c) An electron-doped metallic state.
}
\label{TOfig1}
\end{figure}

When the band is half-filled with one electron 
per site on average($n=1$), each electron tends to be localized on 
a separate lattice site and the spin tends to 
be antiferromagnetically correlated.  When $U/t$ is large enough, 
the groundstate is insulating, which is 
called Mott's insulator (Fig.\ref{TOfig1}(a)), 
and the groundstate is separated from the charge excited states
with a many-body energy gap --- Mott gap.
When we inject carriers (usually with a chemical doping 
by adding or replacing to other elements) to increase (electron doping) or decrease (hole doping) the filling from unity, the Mott gap collapses 
for large enough doping, and the 
system becomes metallic. This is the metal-insulator transition or
the Mott transition, which is widely observed in 
strongly correlated materials.
In these materials, a state occupied simultaneously 
by up an down-spin electrons  --- which we call a doublon --- 
and holes carry the current.
After the discovery of the high-temperature superconductivity 
in cuprates, carrier-doped Mott insulators
have been subject of a huge number of experimental and theoretical studies.

\begin{figure}[thb]
\centering 
\includegraphics[width=11.cm]{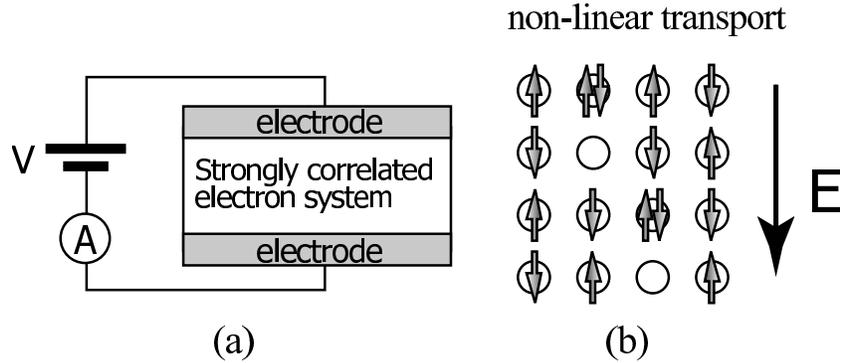}
\caption{(a) Schematic experimental configuration.
(b) Carriers (doublons and holes) created by an 
external electric field.
}
\label{TOfig2}
\end{figure}

Now, let us consider what will happen if we 
attach a set of electrodes to a strongly-correlated 
sample, and apply a large bias voltage across the electrodes
(Fig. \ref{TOfig2} (a)). 
Although the setup may seem simple enough, there is a 
profound physics involved.

In regions near the electrodes (or near the interface 
in the case of heterojunctions between a strongly-correlated 
and ordinary materials), a 
``band bending'' similar to doped semiconductors
can take place and lead to an interface Mott transition 
 when the filling becomes one \cite{OkaNagaosa}. 
The width of the insulating layer changes as the 
applied bias is changed, which dominates the 
behavior of the non-linear transport 
(i.e., the $I-V$ characteristics).  
The result (with DMRG + Hartree potential) 
for the band-bending effects in this case 
can be understood if we assume a 
{\it local equilibrium} for the relation between 
the density of electrons and the potential.  
The local properties are determined by the Hartree potential
governed by Poisson's equation, which in turn 
determines the local chemical potential and 
controls the metal-insulator transition.

Even more interesting, however, is the case 
where we no longer have local equilibrium. Specifically, 
a {\it quantum many-body breakdown of a Mott's insulator} takes place when the 
applied electric field is large enough and creates 
doublons and holes in the Mott insulating groundstate (Fig. \ref{TOfig2} (b)) \cite{Oka2003,Oka2004b,Oka2005a}.

\begin{figure}[thb]
\centering 
\includegraphics[width=11.5cm]{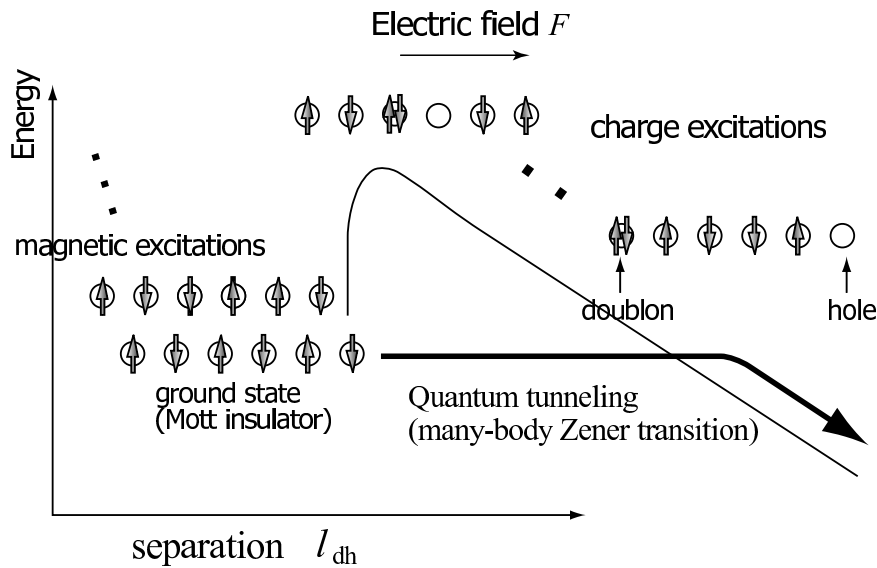}
\caption{Dielectric breakdown of a Mott insulator in 
a strong electric field due to many-body Landau-Zener transition: 
The groundstate and excited states
with charge excitations are separated by an
energy barrier, and quantum tunneling among 
many-body states takes place when the 
electric field is strong enough.
}
\label{fig:energydiagram}
\end{figure}

The creation mechanism is a many-body analog of 
the ``Zener breakdown", well-known in semiconductor physics
\cite{Zener}.  
Namely, while we cannot use the notion of the electronic 
band structure for correlated electron systems, 
we can envisage the carrier-creation process 
as a tunneling across a kind of barrier.  
As displayed in Fig. \ref{fig:energydiagram}, production of 
carriers occurs through tunneling between the Mott insulating groundstate
and excited states with doublons and holes.
If we denote the distance between a doublon and a hole 
by $l_{{\rm dh}}$, the energy profile as a function of $l_{{\rm dh}}$ 
roughly reads
\begin{equation}
\Delta E\sim U-l_{{\rm dh}}F,
\end{equation}
where $F$ is the strength of the electric field.
The profile curve reaches the energy before the 
creation of the doublon-hole pair for the 
separation at which
\begin{equation}
\Delta E\sim U-\bar{l}_{\rm dh}F=0,
\label{eq:delEzero}
\end{equation}
to which the tunneling becomes possible.  
There is a threshold field strength for this process to occur.  
This is because larger 
quantum fluctuations are required to have a larger 
separation between doublon and hole
in the Mott insulator.  

In other words, the overlaps of
many-body wave function of the groundstate 
and excited states decrease rapidly for large $l_{\rm dh}$.  
We can formulate this with the Landau-Zener picture 
in the time-dependent gauge for the 
external electric field, for which, 
as we shall show below\cite{Oka2005a}(eq.(\ref{eq:lzHub})), the threshold 
field strength is given by
\begin{equation}
F_{{\rm th}}=\frac{\Delta_{{\rm c}}(U)^2}{8t_{{\rm hop}}},
\end{equation}
where $\Delta_{{\rm c}}(U)$
is the charge gap (i.e., the Mott gap),
and the tunneling rate per length
is given by
\begin{equation}
\Gamma(F)/L=-\frac{2F}{h}a\ln\left[1-p(F)\right],
\end{equation}
where $p(F)=e^{-\pi\frac{F_{{\rm th}}}{F}}$ is the 
tunneling probability and $a$ a non-universal constant
depending on the detail of the system.
The tunneling rate $\Gamma(F)/L$, being 
related to the production rate of carriers, is 
directly related to physical properties in the bulk 
if the interface effect is neglected.  
Indeed, such a non-linearity in the  $I-V$ 
characteristics has been observed in real materials, 
most prominently in a one-dimensional copper oxide \cite{tag}. 

Another interesting consequence of eq.(\ref{eq:delEzero}) 
is that it gives the ``critical separation" of the 
doublon-hole excitation $\bar{l}_{{\rm dh}}=U/F_{{\rm th}}$. 
This has to do with the convex shape of the energy profile 
against  $l_{\rm dh}$ (Fig.\ref{fig:energydiagram}), which is 
reminiscent of the energy profile 
for the  standard nucleation theory 
that treats the critical size of a stable-phase
droplet to grow without being crushed, 
although the physics involved is quite different. 
In the present case, when
the field is greater than the threshold, the
{\it electric-field induced metallic state}, where 
doublon-hole pairs continue to be created, becomes 
the stable phase. 

The first goal of this 
article is to derive the relations 
presented above and study the 
creation mechanism of carriers (this part is the 
extended argument of our papers \cite{Oka2003,Oka2005a}). 
We need to treat the process quantum mechanically 
and in a many-body formulation.
In doing so, we present 
a renewed and unified interpretation of the Zener transition
of insulators. 
The key quantity is the effective Lagrangian of quantum dynamics  (see \S \ref{sec:Heisenberg-Euler} for a detailed introduction)
which is define by \cite{Oka2005a}
\begin{equation}
\mathcal{L}(F)=-\frac{i}{L^d}\lim_{t\to \infty}\frac{1}{t}\ln\Xi(t),
\end{equation}
where $\Xi(t)$ is the groundstate-to-groundstate transition amplitude 
and $L^d$ is the volume of the $d$ dimensional system
and $L$ the linear size.
There is a deep relation between the theories of 
dielectric breakdown in condensed matter
and a branch in quantum field theory known as non-linear quantum
electrodynamics (QED) (table: \ref{fig:theories}). 
The effective Lagrangian defined above 
coincides with the Heisenberg-Euler effective Lagrangian for 
non-adiabatic evolution \cite{Heisenberg1936,Oka2005a}. 
The effective Lagrangian 
have been used to study the Schwinger mechanism
of electron-positron pair production from the QED 
vacuum in strong electric fields \cite{Schwinger1951}.
In fact, we show that the Schwinger mechanism 
and the Zener tunneling are equivalent, where
the effective action coincides
if we consider the breakdown of  simple Dirac type band insulators.
Furthermore, the effective action gives 
the non-adiabatic extension of the Berry phase theory 
of polarization.

\begin{table}[htbp]
\begin{tabular}{l|c|c}
\hline
&
dielectric breakdown in cond. matter&non-linear QED\\
\hline
\hline
mechanism&Zener breakdown\cite{Oka2003}&Schwinger mechanism \cite{Schwinger1951}\\
excitation&
electron (doublon)-hole pair&electron-positron pair\\
effect of interaction&many-body Landau-Zener &back reaction\\
non-linear polarization&cross correlation (ME effect)
&photon-photon interaction\\
&non-adiabatic Berry phase theory \cite{Oka2005a}&-\\
\hline
\end{tabular}
\caption{Relation between the theory of dielectric breakdown in condensed matter
and non-linear QED from the point of
view of the effective Lagrangian.}
\label{fig:theories}
\end{table}

\begin{figure}[thb]
\centering 
\includegraphics[width=11.5cm]{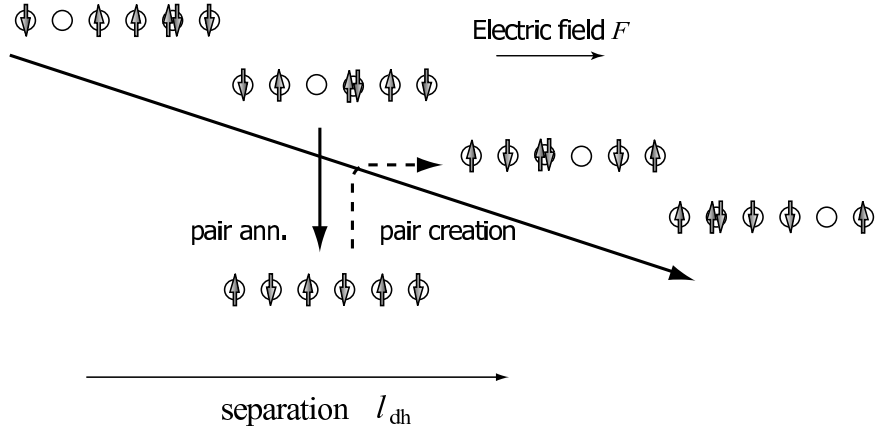}
\caption{Annihilation processes for carriers 
in a correlated electron system. 
}
\label{fig:annihilation}
\end{figure}

In the latter part of the article we shall 
discuss the effect of annihilation of doublon-hole pairs 
(Fig. \ref{fig:annihilation}).
In a one-dimensional system, a doublon and a hole 
cannot pass each other without 
being pair-annihilated even as virtual processes.  
Since the groundstate is locally stable, 
the many-body state tends to remain in the ground state, but 
there should be a finite probability for the 
state to ``branch into" excited states 
through many paths in the many-body energy space. 
Thus, the long-time behavior of the wave function 
involves numerous scattering processes in the
energy space, where, as we shall see, the phase interference 
plays a key role.
We can indeed regard the phase before 
the dielectric breakdown takes place 
as a dynamical localization in the many-body 
energy space, which reduces the tunneling rate and 
makes the groundstate survive\cite{Oka2004b,Oka2004a}.
A statistical mechanical treatment helps 
in understanding this, and we briefly discuss it 
in terms of the quantum walk.

A brief comment on the numerical methods used in the article: 
In order to understand the non-equilibrium processes, 
we need to integrate the time-dependent, many-body Schr\"odinger equation 
to look at the evolution of the many-body wave function 
in, say, the Hubbard model in strong electric fields.  
This is a formidable task, for which no analytically exact treatment  
is known, so that we rely on several numerical methods, which include 
the exact diagonalization and the time-dependent 
density matrix renormalization group method 
(td-DMRG; The version of td-DMRG we adopt is
the one proposed by White and Feiguin\cite{White2004}).

\section{Non-adiabatic evolution and pair creation of carriers}

\subsection{Electric fields and gauge transformation}

When we describe a system in 
finite electric fields, we can choose from 
two gauges.  One is the case where we have a 
slanted electrostatic potential, with the 
gauge field $A^\mu=(Fx,0)$ 
for a one-dimensional system with $F=eE$ being the 
electric field, while the other 
represents the electric field via a 
time-dependent vector potential, $A^\mu=(0,-Ft)$.

\begin{figure}[h]
\centering 
\includegraphics[width= 10cm]{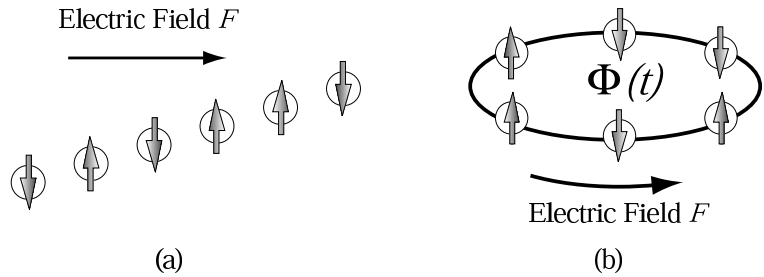}
\caption{(a) Time-independent gauge. (b) Time-dependent gauge.
}
\label{fig:twogauges}
\end{figure}

In the first gauge, the tilted potential enters in the Hamiltonian as
\begin{equation}
H(F)=H_0+F\hat{X},\quad \hat{X}=\sum_jjn_j,
\end{equation}
where $\hat{X}$ is the position operator\cite{Resta1998} and $H_0$ the original Hamiltonian.
This gauge is {\it in}compatible with 
systems with periodic boundary 
conditions. 
The Hamiltonian $H(F)$ 
in fact becomes an unbounded operator in an infinite system, 
since one can lower the energy indefinitely by moving an electron 
to $j\to -\infty$.

In the other gauge, which we call the time-dependent gauge, 
the hopping term of the Hamiltonian becomes time-dependent as
\begin{equation}
H(\phi(t))=-t_{{\rm hop}}\sum_{i\sigma}
\left(e^{i\frac{2\pi}{L}\phi(t)}c^\dagger_{i+1\sigma}c_{i\sigma}+\mbox{h.c}\right)
+\hat{V},
\label{2HamiltoniangeneralTD}
\end{equation}
where $\Phi(t)$ represents a time-dependent Aharonov-Bohm(AB) flux,
\begin{equation}
\phi(t)\equiv \Phi(t)/\Phi_0=FLt/h.
\label{2timeperiod}
\end{equation}
Physically, this gauge amounts to 
considering a periodic system (a ring) and a 
magnetic flux piercing the ring, where 
the time-dependent flux induces electric fields 
by Faraday's law  (Fig. \ref{fig:twogauges}). 
For a higher-dimensional system, the 
ring becomes a (generalized) torus. 
The time-dependent gauge is 
suited for periodic systems 
since it is compatible with the 
lattice translation symmetry. 
The electric current operator 
is obtained by differentiating 
the Hamiltonian by $A^1$ as 
\begin{equation}
J(\phi)=-\frac{dH(\phi)}{dA^1}=-it_{{\rm hop}}\sum_{i\sigma}
\left(e^{i\frac{2\pi}{L}\phi}c^\dagger_{i+1\sigma}c_{i\sigma}
-e^{-i\frac{2\pi}{L}\phi}c^\dagger_{-i\sigma}c_{i+1\sigma}\right).
\end{equation}
There exists an important operator relation
among $H$, $J$ and $\hat{X}$, 
\begin{equation}
J(\phi)=\frac{i}{\hbar}\left[H(\phi),\hat{X}\right], 
\end{equation}
which comes from Heisenberg's equation of motion for 
the current operator, $J(\phi)=\frac{d}{dt}\hat{X}$.

We can relate the two gauges with 
a twist operator\cite{Avron1992} defined by
\begin{equation}
g(\phi)=e^{-i\frac{2\pi}{L}\phi\hat{X}},
\end{equation}
and the two Hamiltonians are related by a gauge transformation
generated by the twist operator, i.e.,
\begin{equation}
H(F)=g^\dagger(\phi(t))H(\phi(t))g(\phi(t))-i g^\dagger (\phi(t))\pa_t g(\phi(t)).
\label{2gaugetransformation}
\end{equation}

\subsection{Heisenberg-Euler effective Lagrangian}
\label{sec:Heisenberg-Euler}
We first discuss the non-adiabatic evolution of electron 
wave functions in 
insulators (either one-body or many-body) in strong electric fields.
Let us consider an insulator at $T=0$ and $F=0$, which is 
described by the groundstate wave function $|\Psi_0\ket$.
We then switch on the electric fields at $t=0$ 
to study the quantum mechanical evolution of the system.
We limit our discussions to coherent dynamics 
and ignore the effect of dissipation due to 
heat bath degrees of freedom as well as 
boundary effects near the electrodes. 

A key quantity to study the non-adiabatic evolution
and quantum tunneling in strong electric fields 
is the (condensed-matter counterpart to the) 
effective Lagrangian introduced for QED by Heisenberg 
and Euler\cite{Heisenberg1936}.
In the time-independent gauge, the electrons are described by the solution 
of the Schr\"odinger equation,
\begin{equation}
|\Psi(t)\ket=e^{-itH(F)}|\Psi_0\ket,
\end{equation}
where we have put $\hbar=1$.  The overlap of the solution 
with the groundstate for $F=0$ --- groundstate-to-groundstate
transition amplitude --- should contain the 
information on the tunneling processes, so we define
\begin{equation}
\Xi(t)=\bra \Psi_0|e^{-itH(F)}|\Psi_0\ket e^{itE_0},
\end{equation}
where we have factored out the trivial 
dynamical phase of the groundstate, 
$E_0=\bra\Psi|H(F=0)|\Psi\ket$.
In the case of the time-dependent gauge, we need to be careful, since the
groundstate is $\phi$ dependent.
If we denote $|0;\phi\ket$ as the instantaneous 
groundstate of $H(\phi)$, the
groundstate-to-groundstate transition amplitude
becomes
\begin{eqnarray}
\Xi (\tau)
=\bra 0;\phi(\tau)|\hat{T}e^{-\frac{i}{\hbar}
\int_0^\tau H(\phi(s))ds}|0;\phi(0)\ket e^{\frac{i}{\hbar}\int_0^\tau E_0(\phi(s))ds},
\label{ggamplitude}
\end{eqnarray}
where $\hat{T}$ stands for the time ordering, and 
$ E_0(\phi)=\bra 0;\phi(\tau)|H(\phi)|0;\phi\ket$ 
the dynamical phase of the groundstate.

\begin{figure}[thb]
\centering 
\includegraphics[width=3.5cm]{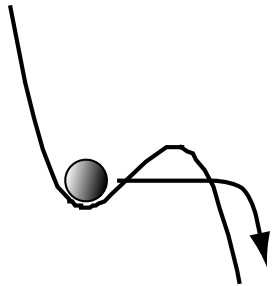}
\caption{The original problem studied by 
Callan and Coleman in which quantum tunneling 
from an unstable vacuum is considered \cite{CallanColeman1977}.
}
\label{fig:breakdown}
\end{figure}
We define\cite{Oka2005a} the effective Lagrangian by 
\begin{equation}
\mathcal{L}(F)=-\frac{i}{L^d}\lim_{t\to \infty}\frac{1}{t}\ln\Xi(t),
\end{equation}
where $L^d$ is the volume of the $d$ dimensional system with 
a linear dimension of $L$.  
We can also regard the Lagrangian as the 
exponent of the asymptotic behavior of the amplitude, 
$\Xi (\tau)\sim e^{i\tau L^d\mathcal{L}(F)}$. 
When the electric field is large enough, the groundstate 
becomes unstable with the quantum tunneling to excited states 
activated. 
The tunneling rate is described by the 
imaginary part of the effective Lagrangian,
\begin{equation}
\Gamma(F)/L^d \equiv 2\mbox{Im}\;\mathcal{L}(F),
\end{equation}
which gives the rate of the exponential decay 
of the vacuum (groundstate).  In the quantum field theory, 
the decay rate of an unstable vacuum has been 
discussed by Callan and Coleman, 
where the tunneling takes place when the 
potential is suddenly changed by an external 
field\cite{CallanColeman1977} (Fig. \ref{fig:breakdown}).

\begin{figure}[h]
\centering 
\includegraphics[width= 12cm]{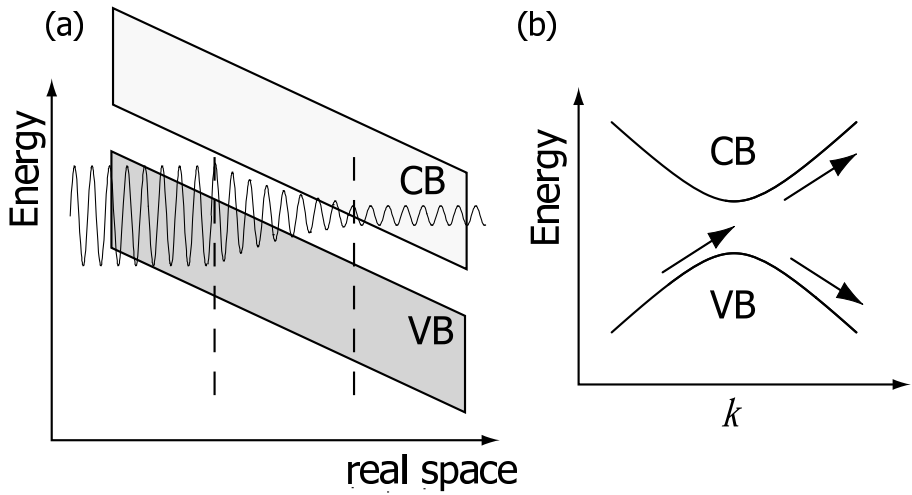}
\caption{Two models of the dielectric breakdown studied by
Zener. (a) Time-independent gauge. (b) Time-dependent gauge. 
}
\label{fig:zener}
\end{figure}

As we shall see later in several models, 
in the theory of dielectric breakdown, 
the tunneling corresponds to creation of
charge carriers. In band insulators the 
carriers are electrons and holes, while 
in Mott insulators they are doublons and holes.
If we neglect boundary effects and assume that all the 
carriers are absorbed by electrodes, 
we can conclude that the tunneling rate 
is proportional to the leakage current, i.e.,
\begin{equation}
J_{{\rm leak}}\propto \Gamma(F)/L^d.
\end{equation}
Indeed, this is the original picture of 
Zener when he calculated the leakage current in 
a simple band insulator \ref{fig:zener}.
Zener has studied the dielectric breakdown
in a simple one-dimensional insulator
using the time-independent gauge \cite{Zener1934} as well as the
time-dependent gauge \cite{Zener}.
In the former, he has calculated the 
tunneling probability of Bloch functions
in constant electric fields to obtain the tunneling rate.
In the time-dependent gauge, he has described the problem as 
a system with a time-dependent Hamiltonian represented by
a two-by-two matrix to study the 
tunneling near the level anti-crossing,
which is now known as the Landau-Zener transition \cite{Landau,Zener}.

The reason why we have called $\mathcal{L}(F)$ the 
effective Lagrangian is that it coincides with the 
Heisenberg-Euler effective Lagrangian in QED \cite{Heisenberg1936}. 
They have studied the dynamics and nonlinear responses of 
the QED vacuum in strong electric fields 
by calculating the effective Lagrangian 
(for a review, see e.g. \cite{Dittrich}). 
By integrating out high-energy degrees of freedom 
(polarization processes due to electron-positron creation/annihilation)
they arrived at an effective description of the 
low-energy degrees of freedom, namely the quantum correction, 
originating from the fluctuation of the QED vacuum, 
to the Maxwell theory of electromagnetism. 
Indeed, if we apply our formalism 
to band insulators with Dirac-type (mass-gapped) dispersions, 
the effective Lagrangian coincides with the 
Heisenberg-Euler Lagrangian with some 
modifications coming from the 
Brillouin zone structure of the Bloch waves as will be shown in the 
next section.  The correspondence between the 
two phenomena is straightforward: The ground-state of the 
insulator translates to the QED vacuum, 
charge excitations to the electron-positron pairs.
The tunneling rate also has its QED counterpart, namely the
vacuum decay rate due to the Schwinger mechanism --- 
creation of electron-positron pairs in strong electric 
fields \cite{Schwinger1951}. 

\subsubsection{Related theories}
We conclude this section with comments on the relation of the 
effective Lagrangian approach 
to earlier theoretical frameworks.
\paragraph{Berry's phase theory of polarization:}
In the Berry's phase 
theory of polarization\cite{Resta1992,KingSmith1993,Resta1998,Resta1999,Nakamura2002}, 
the ground-state expectation value of the twist operator 
$e^{-i\frac{2\pi}{L}\hat{X}}$, which shifts the phase of electron wave 
functions 
on site $j$ by $-\frac{2\pi}{L}j$ \cite{Nakamura2002}, plays a crucial role. 
It was revealed that the real part of a quantity 
\begin{equation}
w=\frac{-i}{2\pi}\ln\bra 0|e^{-i\frac{2\pi}{L}\hat{X}}|0\ket
\end{equation}
gives the linear-response electric polarization, 
$P_{\rm el}=-\mbox{Re}w$ \cite{Resta1998},
while its imaginary part gives a criterion for metal-insulator
transition, i.e.,
$D=4\pi\mbox{Im}w$ is finite in insulators 
and divergent in metals \cite{Resta1999}.
The present effective action is regarded as 
a non-adiabatic (finite electric field) extension of $w$.
To give a more accurate argument, 
recall that the effective Lagrangian can 
be expressed as
\begin{equation}
\mathcal{L}(F)\sim \frac{-i\hbar}{\tau L}\ln\left(
\bra 0|e^{-\frac{i}{\hbar}\tau(H+F\hat{X})}
|0\ket e^{\frac{i}{\hbar}\tau E_0}\right)
\end{equation}
for $d=1$. 
Let us set $\tau=h/LF$ and consider the small $F$ limit.  
For insulators we can replace $H$ with the
groundstate energy $E_0$ to have 
$\mathcal{L}(F)\sim wF$ 
in the linear-response regime. Thus the real part of 
Heisenberg-Euler's expression\cite{Heisenberg1936} for 
the non-linear polarization 
$P_{\rm HE}(F)=-\pa \mathcal{L}(F)/\pa F$ 
naturally reduces to the Berry's phase 
formula $P_{\rm el}$ in the $F\to 0$ limit (cf. eq.(\ref{2Schwinger2}) below).
Its imaginary part, which is related to 
the decay rate as $\mbox{Im}P_{\rm HE}(F)=-\frac{\hbar}{2}
\frac{\pa \Gamma(F)/L}{\pa F}$, reduces to $ -D/4\pi$ 
and gives the criterion for the transition, originally proposed
for the zero field case.

\paragraph{Non-Hermitian quantum mechanics}
The dielectric breakdown of Mott insulators was also studied in the
framework of non-Hermitian quantum mechanics\cite{Fukui,NakamuraHatano06}.
Fukui and Kawakami studied a non-Hermitian Hubbard model in
which the leftward and rightward hopping integral are assumed to be unequal \cite{Fukui}. 
The non-Hermiticity is assumed to represent the coupling
of the system with a ``dissipative environment".
With the Bethe ansatz solution they have observed the gap between the 
groundstate and the first excited state to close
when the hopping asymmetry is large enough. 
It seems that the remaining question is to 
relate this result with measurable quantities.

\subsection{Zener breakdown of band insulators revisited --- 
Non-adiabatic geometric phase and the Schwinger mechanism}

\begin{figure}[thb]
\centering 
\includegraphics[width=12.cm]{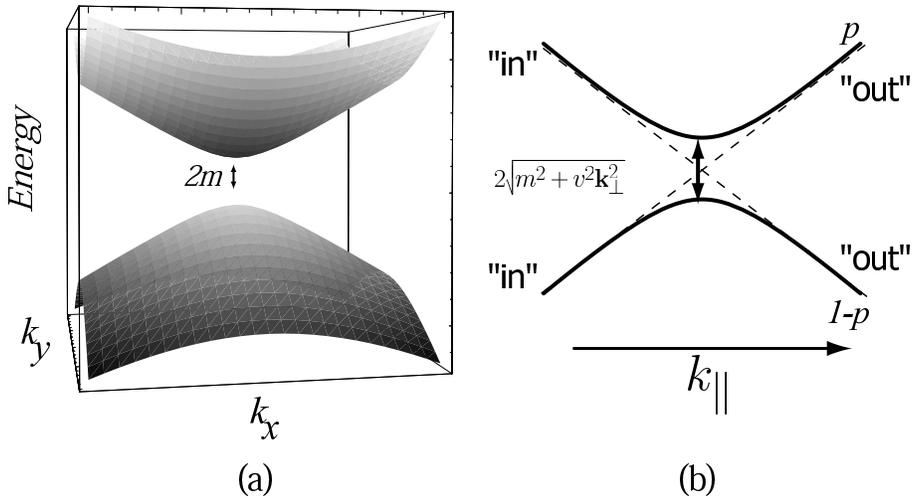}
\caption{(a) Energy levels of a Dirac band in 2D.
(b) A one dimensional slice of the higher dimensional Dirac
band in which carriers (doublons and holes) are created by 
an external electric field in the $k_\parallel$ direction.
}
\label{TOfig8}
\end{figure}
Before examining the dielectric breakdown of correlated 
electron systems, let us first discuss the dielectric breakdown of 
band insulators in an electric field $\Vect{F}$
within the effective-mass picture. 
This will turn out to be heuristic, since 
we can obtain an analytic expression for the effective 
Lagrangian which can be readily applied to general
band insulators. 

For simplicity we take a pair of hyperbolic bands 
$\ve_\pm(\Vect{k})=\pm \sqrt{V^2+v^2k^2}$ 
(considered here in $d$ spatial dimensions), 
where $2V$ is the band gap, $-(+)$ denote the 
valence (conduction) band, 
and $v$ the asymptotic slope of the dispersion.  

We first obtain the groundstate-to-groundstate 
transition amplitude with the time-dependent gauge
in the periodic boundary condition. 
There, a time-dependent AB-flux in units of the 
flux quantum, $\phi(\tau)=FL\tau/h$ 
(with the electronic charge $e=1$ and 
$L$ being the system size), is introduced to 
induce an electric field $F$, which makes the Hamiltonian 
time dependent as 
\begin{equation}
H(\phi(\tau))=
\sum_{\bf{k},\alpha=\pm}\ve_\alpha\left(\Vect{k}+\frac{2\pi}{L}\phi(\tau)\Vect{e}_{\parallel}\right)
c_{\alpha}^\dagger(\Vect{k})c_{\alpha}(\Vect{k}).
\end{equation}
Here $\Vect{e}_{\parallel}$ is the 
unit vector parallel to $\Vect{F}$, 
and $c_{\alpha}^\dagger(\Vect{k})$ the creation operator with 
spin indices dropped. 
If we denote the ground state of $H(\phi)$ as $|0;\phi\ket$ 
and its energy as $E_0(\phi)$, the groundstate-to-groundstate 
transition amplitude reads
\begin{eqnarray}
\Xi (\tau)
=\bra 0;\phi(\tau)|\hat{T}e^{-\frac{i}{\hbar}
\int_0^\tau H(\phi(s))ds}|0;\phi(0)\ket e^{\frac{i}{\hbar}\int_0^\tau E_0(\phi(s))ds},
\label{ggamplitude}
\end{eqnarray}
where $\hat{T}$ stands for the time ordering.
The effective Lagrangian $\mathcal{L}(F)$ for 
the quantum dynamics is defined from the asymptotic behavior, 
$\Xi (\tau)\sim e^{\frac{i}{\hbar} \tau L^d\mathcal{L}(F)}$.

The dynamics of the one-body model can be solved analytically (Fig.\ref{TOfig8} (b)), 
since we can cut the dispersion in $d$ spatial dimensions 
into slices, each of which 
reduces to Landau-Zener's two band model in 1D\cite{Landau,Zener}.
Namely, if we decompose the $k$ vector
as $(\Vect{k}_\perp,k_\parallel)$, 
where $\Vect{k}_\perp$ ($k_\parallel$) is the 
component perpendicular (parallel) 
to $\Vect{F}$, where each slice for a given $\Vect{k}_\perp$ 
is a copy of Landau-Zener's model 
with a gap $\Delta_{\rm band}(\Vect{k})\equiv 2\sqrt{V^2+v^2k_\perp^2}$.
The Landau-Zener transition takes place around the level anti-crossing 
for which $k_\parallel+\frac{2\pi}{L}\phi(\tau )$ moves across the 
Brillouin zone(BZ) 
in a time interval $\delta \tau=h/F$.  The process can be expressed 
as a scattering and the Bogolubov coefficients 
between the ``in" and ``out" states (see Fig.\ref{TOfig8}) is given by the 
solution to the two band problem, i.e.,
\begin{eqnarray}
c_+^\dagger(\Vect{k})\to 
\sqrt{1-p(\Vect{k})}e^{-i\chi(\Vect{k})}c_
+^\dagger(\Vect{k})+\sqrt{p(\Vect{k})}c_-^\dagger(\Vect{k}),\nonumber\\
c_-^\dagger(\Vect{k})\to -\sqrt{p(\Vect{k})}c_+^\dagger(\Vect{k})+
\sqrt{1-p(\Vect{k})}e^{i\chi(\Vect{k})}c_-^\dagger(\Vect{k}).
\label{2Bogoliubov}
\end{eqnarray}
Here the tunneling probability for each $\Vect{k}$ is given by 
the Landau-Zener(LZ) formula\cite{Landau,Zener}, 
\begin{equation}
p(\Vect{k})=\exp\left[-\pi
\frac{(\Delta_{\rm band}(\Vect{k})/2)^2}{vF} \right].
\label{LZS1}
\end{equation}
On the other hand, 
the phase $\chi(\Vect{k})=-\theta(\Vect{k})+\gamma(\Vect{k})$ 
appearing in the Bogolubov coefficients 
consists of the trivial dynamical phase,
$
\hbar \theta(\Vect{k})=
\int_0^{\delta \tau}\ve_{+}(\Vect{k}+
\frac{2\pi}{L}\phi(s)\Vect{e}_\parallel) ds,
$
and the Stokes phase\cite{Zener,Kayanuma1993},
\begin{eqnarray}
\gamma(\Vect{k})=\frac{1}{2}\mbox{Im}\;\int_0^\infty 
ds\frac{e^{-i(\Delta_{\rm band}(\Vect{k})/2)^2 s}}{s}
\left[\cot (vFs)-\frac{1}{vFs}\right].
\label{2StokesphaseDDimBand}
\end{eqnarray}
The Stokes phase, a non-adiabatic 
extension of Berry's geometric phase \cite{Berry1984}, 
depends not only on the topology of the path but, 
unlike the adiabatic counterpart, also on the
field strength $F$ \cite{Kayanuma1997}.
In terms of the fermion operators 
the groundstate is obtained by filling the lower band
 $|0;\phi\ket=\prod_{\Vect{k}}
c^\dagger_-(\Vect{k}-\frac{2\pi}{L}\phi\Vect{e}_{\parallel})|{\rm vac}\ket
$, 
where $|{\rm vac}\ket$ is the fermion vacuum with 
$c_{\pm}(\Vect{k})|{\rm vac}\ket =0$.
If we assume that excited charges are absorbed by electrodes
we obtain from eqs.(\ref{ggamplitude}), (\ref{2Bogoliubov})
\begin{eqnarray}
\mbox{Re}\;\mathcal{L}(F) &=&-F
\int_{\rm BZ} \frac{d\Vect{k}}{(2\pi)^{d}}\frac{\gamma
(\Vect{k})}{2\pi},\nonumber\\
\mbox{Im}\;\mathcal{L}(F) &=&-F\int_{\rm BZ} 
\frac{d\Vect{k}}{(2\pi)^{d}}\frac{1}{4\pi}\ln \left[ 1-p(\Vect{k})\right],
\label{2Schwinger2}
\end{eqnarray}
where the dynamical phase $\theta$ cancels the 
factor $e^{\frac{i}{\hbar}\int_0^\tau E_0(\phi(s))ds}$ 
in eq.(\ref{ggamplitude}).

\begin{figure}[thb]
\centering 
\includegraphics[width=10.cm]{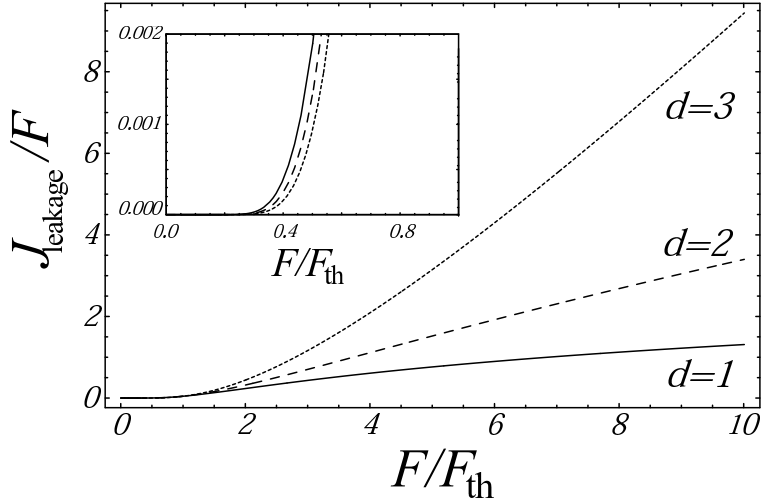}
\caption{The dependence of the conductivity on the 
electric field in the non-linear regime for 
band insulators with spatial dimension $d=1,2,3$.
The inset zooms in the threshold region. 
}
\label{fig:conductivity}
\end{figure}

Integration over $\Vect{k}$ in eq.(\ref{2Schwinger2}) 
leads to the groundstate decay rate per volume for
a $d$-dimensional hyperbolic band,
\begin{eqnarray}
\Gamma(F)/L^d&=&
\frac{F}{(2\pi)^{d-1}h}\left(\frac{F}{v}\right)^{(d-1)/2}\nonumber\\
&&\times
\sum_{n=1}^\infty \frac{1}{n^{(d+1)/2}}
e^{-\pi n\frac{V^2}{vF}}
\left[
\mbox{erf}\left(\sqrt{\frac{nv\pi^3}{F}}\right)\right]^{d-1}.
\label{BandSchwingerFormula}
\end{eqnarray}
The threshold for the tunneling is governed by the 
most nonlinear (actually essentially singular) factor 
in the above formula, namely $e^{-\pi n\frac{V^2}{vF}}$, 
so that the threshold electric field is given by
\begin{equation}
F_{{\rm th}}=\frac{V^2}{v}.
\end{equation}
Although an analytic integration eq.(\ref{2Schwinger2}) is possible 
for a Dirac band (= hyperbolic valence and conduction bands), 
the expression is valid for general band dispersions. 
In fig. \ref{fig:conductivity}, the leakage 
current divided by the field strength, which is proportional 
to $\Gamma(F)/FL^d$, is plotted for the spatial dimension $d=1,2,3$. 
The $F$-dependence is essentially
\begin{equation}
J_{{\rm leakage}}\propto F^{(d+1)/2}
e^{-\pi \frac{F_{{\rm th}}}{F}},
\end{equation}
which has a threshold behavior as shown in the inset of the figure. 
Above the threshold, two regimes exist.
In the medium field regime, the current scales as
$j_{{\rm leakage}}\sim F^{(d+1)/2}$ 
where the power depends on $d$. 
However, when the field strength is 
even stronger, the error function 
appearing in eq.(\ref{BandSchwingerFormula}), 
which is due to the lattice structure (with the $\Vect{k}$
integral restricted to the BZ), starts to take an 
asymptote (erf$(x) \sim (2/\sqrt{\pi})x$).  
Then various factors (including a power of $F$) 
cancel with each other, and 
the leakage current in the $F\to \infty$ limit approaches a {
\it universal} function, 
\begin{equation}
J_{{\rm leakage}}\propto
\Gamma(F)/L^d\to -\frac{F}{h}\ln
\left[1-\exp\left(-\pi \frac{F_{\rm th}}{F} \right)\right],
\label{eq:leakagecurrent}
\end{equation} 
where the $d$ dependence disappears up to a trivial 
$d$-dependent numerical factor. 
This prediction on the non-linear transport can be checked 
experimentally including low-dimensional systems such as 
carbon nanotubes ($d=1$).  
Graphene ($d=2$) is also interesting, but 
this system has a massless Dirac dispersion, 
so that a special treatment should be required.

\paragraph{Comparison to Heisenberg-Euler-Schwinger's 
results in  non-linear QED}

Let us have a closer look at the decay of the QHE vacuum.  
In 1936, Heisenberg and Euler studied Dirac particles in 
strong electric fields, and 
discussed non-linear optical responses 
of the QED vacuum --- vacuum polarization --- 
in terms of an effective Lagrangian \cite{Heisenberg1936}. 
Later, Schwinger refined their approach and calculated the 
vacuum decay rate \cite{Schwinger1951}\footnote{
For references on the 
effective-action approach of non-linear 
electrodynamics, see \cite{ItzyksonZuber,Dittrich}.}.
Up to the one-loop level, Schwinger calculated the 
vacuum-to-vacuum transition amplitude 
using the proper time regularization method to obtain
\begin{equation}
\Delta\mathcal{L}^{{\rm QED}}(F)=\frac{1}{8\pi^2}\int_0^\infty\frac{ds}{s^2}
\left[F\cot(Fs)-\frac{1}{s}\right]e^{-ism_e^2}
\label{eq:HES}
\end{equation}
for (3+1)-dimensional QED, where $m_e$ is the electron mass.
The integrand has a pole in the complex domain 
and has an imaginary part, which gives
\begin{equation}
\Gamma(F)^{{\rm QED}}/L^d=\frac{\alpha F^2}{2\pi^2}\sum_{n=1}^\infty
\frac{1}{n^2}\exp\left(-\frac{n\pi m_e^2}{|F|}\right),
\label{eq:qedschwinger}
\end{equation}
the famous Schwinger's formula for the 
electron-positron pair creation rate \cite{Schwinger1951}, where 
$\alpha=1/137$ is the fine-structure constant. 

Thus the expression for the QED
effective Lagrangian, eq.(\ref{eq:HES}), 
coincides with the Stokes phase for 
the non-adiabatic Landau-Zener tunneling, 
except for a difference in the momentum integral.  
As we have mentioned above, an important difference 
in lattice systems is that the momentum integral 
is limited to the Brilliouin zone, 
and the decay-rate acquires an extra factor 
(compare eq.(\ref{BandSchwingerFormula}) with erf 
with eq.(\ref{eq:qedschwinger})).  
This modification changes the strong field limit
of the leakage current which leads to the 
universal expression (eq.(\ref{eq:leakagecurrent})). 
Another important difference, which is quantitative, appears in the
threshold voltage:
The threshold for band insulators 
$E_{\rm th}^{\rm band}=F_{\rm th}^{\rm band}/e=V^2/vae$ 
($a$: lattice constant) 
is many orders smaller than
the threshold for the QED instability 
$E^{\rm QED}=\frac{m_e^2c^3}{\hbar}\sim 10^{16}\;\mbox{V/cm}$.
For example, if we have an insulator
with parameters $a=10^{-7}\mbox{cm},\;v=2t_{{\rm hop}}=1\mbox{eV},\;
V=1\mbox{eV}$, 
then the threshold becomes as small as 
$E_{\rm th}^{\rm band}=10^7 \mbox{V/cm}$.

Heisenberg and Euler's original aim 
was to discuss non-linear optical properties
of the vacuum in terms of $\Delta\mathcal{L}$. 
In fact, they calculated the 
effective Lagrangian in the presence of both electric 
and magnetic fields \cite{Heisenberg1936}, 
and obtained
\begin{eqnarray}
\Delta\mathcal{L}^{{\rm QED}}(F)=C\frac{\Vect{E}^2-\Vect{B}^2}{2}+\frac{2\alpha^2}{45m_e^4}
\left[(\Vect{E}^2-\Vect{B}^2)^2+7(\Vect{E}\cdot\Vect{B})^2\right]+\ldots,
\label{eq:qedeffectivelagrangianperturbation}
\end{eqnarray}
where $C$ is a diverging constant that we 
drop after renormalization. 
The electric polarization can be 
obtained from the real part of the effective 
action via 
\begin{equation}
\Delta P(F)=\frac{\pa}{\pa F}\Delta \mathcal{L}(F).
\end{equation}
If we plug in eq(\ref{eq:qedeffectivelagrangianperturbation}), 
the non-linear polarization of Dirac particles becomes
\begin{eqnarray}
\Delta P&=&\frac{2\alpha^2}{45m_e^4}\left(
-4B^2E+14B_\parallel^2E+4E^3
\right)+\ldots,\\
&=&\sum_{n=1}^\infty P^{(n)}(\Vect{B})E^n,
\end{eqnarray}
where $B_\parallel$ is the 
component of $\Vect{B}$ parallel
to $\Vect{E}$, and 
$ P^{(n)}(\Vect{B})$ the $n-$th order 
non-linear polarization.  
Thus we can examine nonlinear polarizations and 
cross correlations (a combined effect of $\Vect{E}, \Vect{B}$) 
with the effective Lagrangian, as touched upon 
in Table \ref{fig:theories}.

\subsection{Dielectric breakdown in a Mott insulator ---
many-body Landau-Zener transition and a nonequilibrium phase diagram}
\begin{figure}[thb]
\centering 
\includegraphics[width= 10cm]{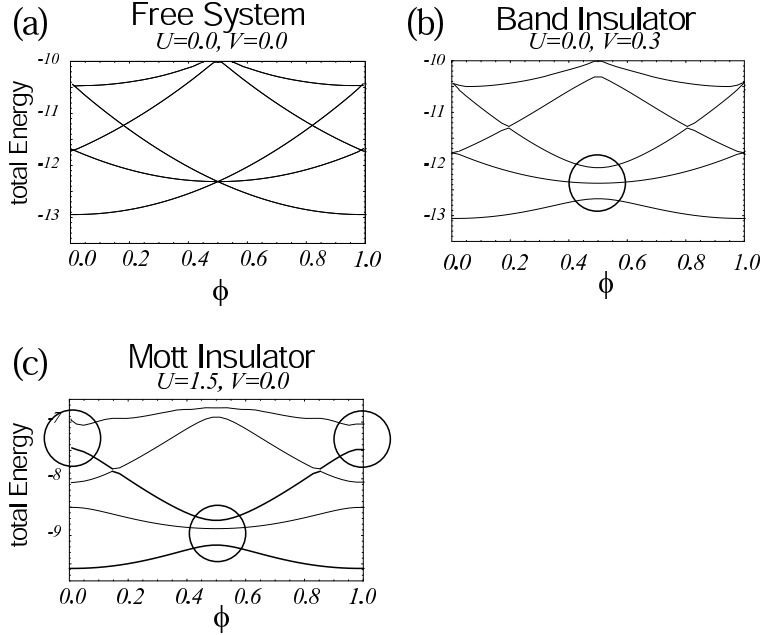}
\caption{
Adiabatic spectrum $E_n(\phi)$ for a 
finite system ($L=10$ here) 
obtained by the Lanczos method.
We plot low-lying excitations in the half-filled 
subspace $N_\up=N_\dw=L/2$.
(a) a noninteracting system in a free space, 
(b) a band insulator ($U/t_{{\rm hop}}=0,V/t_{{\rm hop}}=0.3$), 
and (c) a Mott insulator ($U/t_{{\rm hop}}=1.5, V/t_{{\rm hop}}=0$).
The circles indicate avoided level crossings.
}
\label{fig:adiabaticLevels}
\end{figure}

Before applying the effective Lagrangian approach to
the dielectric breakdown of Mott insulators, 
we need to examine the excitation spectra, which is displayed  
in fig.\ref{fig:adiabaticLevels}.  There we plot, 
for the half-filling, the 
many-body energy levels of the Hamiltonian,
\begin{eqnarray}
H(\phi(t))=-t_{{\rm hop}}\sum_{i\sigma}
\left(e^{i\frac{2\pi}{L}\phi(t)}c^\dagger_{i+1\sigma}
c_{i\sigma}+\mbox{h.c.}\right)+U\sum_in_{i\up}n_{i\dw}+V\sum_i(-1)^in_i.
\label{eq:phiHubbard}
\end{eqnarray}
Here $U$ is the Hubbard repulsion, $V$ a staggered potential 
to introduce valence and conduction bands, so that 
$U=V=0$ 
corresponds to a noninteracting system in a free space,
 $U=0,\;V\ne 0$ a band insulator,
and a large $U$ and $V=0$ a Mott insulator.
In the figure, we have 
only plotted charge excitations (where the charge rapidities 
are excited in the language of Bethe-ansatz solution).  
As can been seen, levels cross in the free model 
while in the band and the Mott insulators 
an energy gap separates the ground state 
from excited states.
The gap is $2V$ for the band insulator.
The Hubbard Hamiltonian eq.(\ref{eq:phiHubbard}) 
with $U\ne 0,\;V=0$ is also exactly solvable in 1D. 
Woynarovich used the Bethe ansatz method 
\cite{Woynarovich1982,Woynarovich19822} 
to study the ground state as well as the excited states 
(see also \cite{Sut,Kabe,Ari}).  
The two solid lines in Mott insulator's spectrum
correspond to the ground state
and the charge-excited state with one pair of complex charge rapidities, 
quantum numbers appearing in the Bethe-ansatz solution.
The energy gap $\Delta E(U)$ between these states 
are known to converge to the Mott gap $\Delta_{\rm Mott}(U)$
in the limit of infinite system,
\begin{equation}
\Delta E(U)\to\Delta_{\rm c}(U).
\end{equation}
An important feature in the spectrum of the 
Mott insulator is that level repulsion 
occurs at many places over the excited states. 
The repulsion comes from Umklapp electron-electron scattering, 
i.e., a scattering process in which the 
momentum sum changes by 
reciprocal lattice vectors.
In band insulators level repulsions obviously 
come from one-body scattering as we have seen above.

Why have we first looked at the adiabatic spectrum? 
There is an important relation between the 
adiabatic energy and the current expectation value.
From the 
Hellmann-Feynman theorem, i.e., 
$\displaystyle{\frac{dE}{d\lambda}=
\frac{\bra\Psi|\pa H/\pa\lambda|\Psi\ket}{\bra\Psi|\Psi\ket}}$ for 
$H(\lambda)|\Psi(\lambda)\ket=E(\lambda)|\Psi(\lambda)\ket$, 
we obtain
\begin{eqnarray}
J_n(\phi)&=&\bra n;\phi|J(\phi)|n;\phi\ket\nonumber\\
&=&\left(\frac{L}{2\pi}\right)\frac{\pa E_n(\phi)}{\pa\phi}\,
\end{eqnarray}
which is valid for all $\phi$. 
If we expand it around $\phi=0$, we get 
\begin{equation}
J_n(\phi)=J_n(0)+\left(\frac{L}{2\pi}\right)\frac{\pa^2 E_n(0)}{\pa\phi^2}\phi
+O(\phi^2).
\end{equation}
Using $\phi=FLt/h$ and defining 
the transport coefficients
$\mathcal{D}_n$ by 
$J_n(\phi)=J_n(0)+\mathcal{D}_n Ft+O(F^2)$, 
we obtain 
\begin{equation}
\mathcal{D}_n(L) =\left(\frac{L}{2\pi}\right)^2\frac{\pa^2 E_n(0)}{\pa\phi^2}.
\end{equation}
When we focus on a dissipationless adiabatic transport at $T=0$, 
the total current thus reads
\begin{equation}
\bra J(t)\ket =\mathcal{D}_0(L)Ft,
\end{equation}
which is determined by 
the Drude weight (charge stiffness) $\mathcal{D}_0(L)$. 
As we can see in Fig.\ref{fig:adiabaticLevels}, even for insulators ((b) and (c)), the Drude weight $\mathcal{D}_0(L)$ of 
a finite system is not necessarily zero.  If we remember 
{\it Kohn's criterion}\cite{Kohn1963} for 
metal-insulator transitions, stated as
\begin{eqnarray}
\lim_{L\to \infty}\mathcal{D}_0(L)=\left\{
	\begin{array}{cl}
	0&\quad\mbox{insulator},\\
	\mbox{finite}&\quad\mbox{perfect metal},
	\end{array}
\right.
\label{2KohnCriterion}
\end{eqnarray}
we can see that we must go to the limit of infinite systems to distinguish 
metals from insulators. Indeed, the problem of taking 
the infinite-size limit is also occurs in the study of dielectric breakdown
in Mott insulators as we shall see later.

\subsubsection{Short-time behavior --- 
an exact diagonalization result}

Since the time evolution of many-body systems cannot be 
treated analytically, we employ numerical methods 
to time-integrate in two steps --- for short-time behavior and 
long-time behavior.  
For the short-time evolution in dielectric breakdown of Mott insulators 
we exactly diagonalize the 
time-dependent Schr\"odinger equation as follows: 
First we start from the ground state of $H(\phi=0)$
at time $t=0$.
The wave function evolves with the phase 
that increases as
\begin{equation}
\phi(t)= 0 \rightarrow FLt/h.
\end{equation}
Here $F=eaE$ is the field strength, 
$L$ the length of the chain.  
We numerically solve the time-dependent Schr\"odinger equation, 
\begin{equation}
i\frac{d}{dt}|\Psi (t)\ket=H(\phi(t))|\Psi (t)\ket.
\end{equation}
We choose the initial state to be the 
ground-state $|0\ket$ of $H(0)$,
which is obtained here by the Lanczos method. 
The time integration of the state vector, which, being a many-body state, 
has a huge dimension, requires a reliable algorithm.  
So we adopt here the Cranck-Nicholson method that guarantees the unitary 
time evolution, where the time evolution is put into a form,
\begin{equation}
|\Psi (t+\Delta t)\ket=e^{ -i\int_t^{t+\Delta t} H(t)}\; dt\;|\Psi (t)\ket\simeq
\frac{1-i\Delta t/2H(t+\Delta t/2)}{1+i\Delta t/2H(t+\Delta t/2)}\;|\Psi (t)\ket,
\end{equation}
which is unitary by definition.  
Here the time step is taken to be small enough ($dt=1.0\times 10^{-2}$ 
with the time in units of $\hbar/t$ hereafter) 
to ensure convergence for $L\leq 10$, 
for which the dimension of the Hamiltonian is $\sim 10^4$.  
We have concentrated on the total $S^z=0$ subspace with 
$N_\uparrow=N_\downarrow=L/2$.

\begin{figure}[htb]
\centering 
\includegraphics[width=10.5cm]{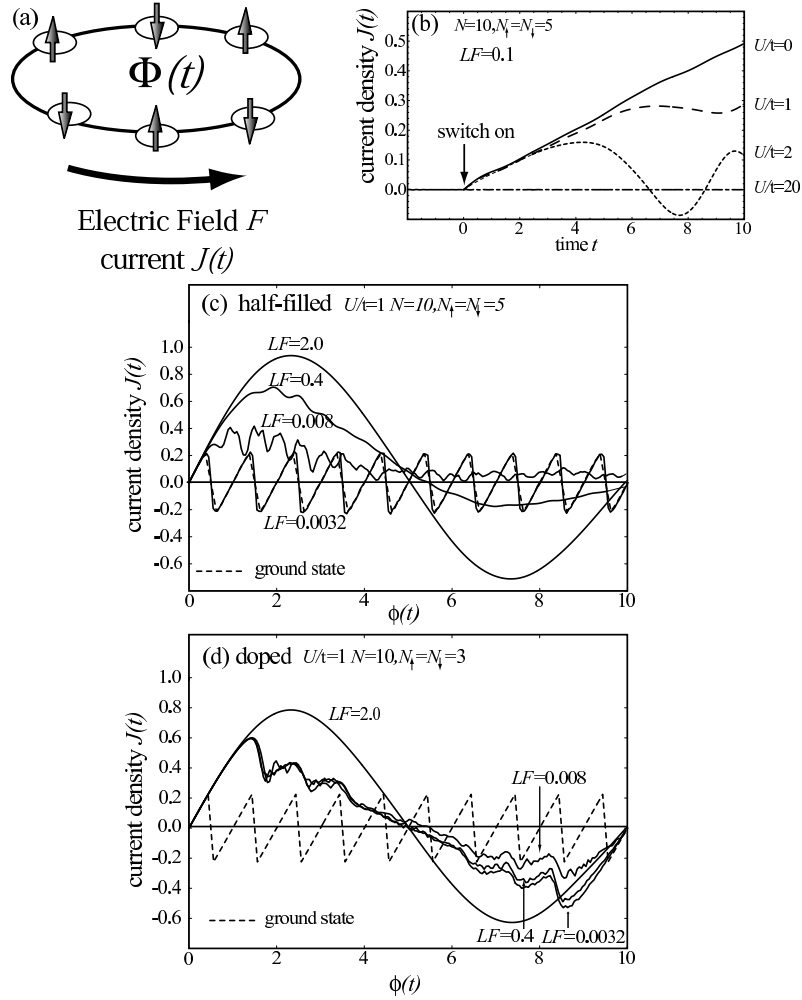}
\caption{
(a) The sample geometry, where 
an AB flux, $\phi(t) = LFt$, increasing linearly with time induces 
an electric force through Faraday's law.
(b) Time evolution of 
the current, $J(t)$, for a half-filled, 10-site 
Hubbard model for 
various strengths of the Hubbard repulsion, $0\leq U/W\leq 5$ 
($W=4t_{\rm hop}$ is the non-interacting band width) 
for a fixed electric field $F=1/10L$.   
Time is measured in units of $\tau_t\equiv \hbar/t_{\rm hop}$, 
$LF$ in $t_{\rm hop}$, and $J(t)$ in 
$1/\tau_t$. The range of the time in this panel is 
restricted to a range of the AB-flux $0\le\phi\le 1$.  
(c) A wider plot of the current 
for various values of $F$ with a fixed 
$U/W=0.25$, again for the half-filled case. 
Here the horizontal axis is $\phi$.  
(d) A plot similar to (c) for a non-half-filled case 
($N_\uparrow=N_\downarrow=3<L/2=5$).
}
\label{currents}
\end{figure}

\paragraph{Evolution of the total current}
We first plot in Fig.\ref{currents}(b) 
the result for the expectation value of the 
current density averaged over the sites, 
$
J=-\frac{it}{L}
\sum_{i,\sigma}\left( e^{i\frac{2\pi}{L}\phi(t)}\;c_{i+1\sigma}^\dagger c_{i\sigma} - {\rm h.c.}
\right). $
The behavior of $J(t)$ for a fixed value of the electric field $F$ 
is seen to fall upon three regimes when $U$ is varied: 
A perfect metallic behavior ($J(t)\propto t$) 
when the electrons are free ($U/W=0$), 
an insulating behavior ($J(t)=0$) when the interaction is strong enough 
($U/W\gg 1$), and 
an intermediate regime of $U/W$ where $J$ is finite 
with some oscillations for finite systems.  
In contrast, a non-half-filled system in nonequilibrium ($F\neq 0$) 
has a time evolution that is 
distinct from the ground-state behavior (Fig.\ref{currents}(d)).  
The difference has its root in the spectral property 
as will be discussed later.  

\begin{figure}[t]
\centering 
\includegraphics[width=9cm]{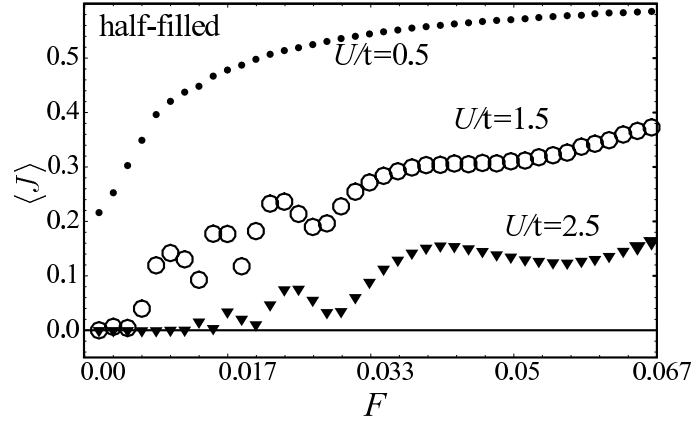}
\caption{Dependence of the averaged current $\langle J(t)\rangle$ 
on $F$ for various 
values of $U/t$ for the half-filled Hubbard model with $L=6$. 
}
\label{IV}
\end{figure}

If we look at the behavior over several periods ($0<\phi<10$) 
for a fixed value of $U/W$ for the half-filled 
(Fig.\ref{currents}(c)) and for a non-half-filled case 
(Fig.\ref{currents}(d)), the result may be summarized as follows: 
\begin{description}
	\item[(i)] Small $F$ regime (Mott insulator preserved at half filling)\\
A drastic difference between the half-filled and doped systems appears 
for small $F$.  When half-filled, $J(t)$ in the limit of 
$F\rightarrow 0$ smoothly approaches a periodic saw-tooth 
behavior with periodicity $\phi=1$, which is 
the AB-oscillation of the ground-state current. 

\item[(ii)] Moderate $F$ regime (metal)\\
In this regime, the current in the half-filled case 
is non-zero and 
shows oscillatory behaviors 
(seen typically in data for $LF=0.008$ in 
Fig.\ref{currents}(c)).

\item[(iii)] Large $F$ regime (perfect metal)\\
When the electric field $F$ becomes large enough, 
the system behaves as a kind of metal.
The current $J(t)$ exhibits a 
long-period ($\Delta\Phi=\Phi_0 L$) oscillation, 
which is the Bloch oscillation, a hallmark of a metal. 
\end{description}
The averaged current, 
\begin{equation}
\langle J\rangle= \frac{1}{T}\int_0^{T}\bra J(t)\ket dt,
\end{equation}
integrated over a quarter of the Bloch period 
(with $\phi(T)=\frac{L}{4}$) 
is plotted against $F$ in Fig.\ref{IV} for various values of $U$.  
We can see that $\langle J \rangle$ becomes nonzero rather 
abruptly at the metallization as $F$ is increased, 
where the threshold electric 
field increases and the $F$-
dependence becomes weaker when we increase $U/t$.  
Just after the metallization 
some oscillation (in the $F$-
dependence this time) is seen for finite systems.

\paragraph{Evolution of the survival probability}
\begin{figure}[h]
\centering 
\includegraphics[width=11.5cm]{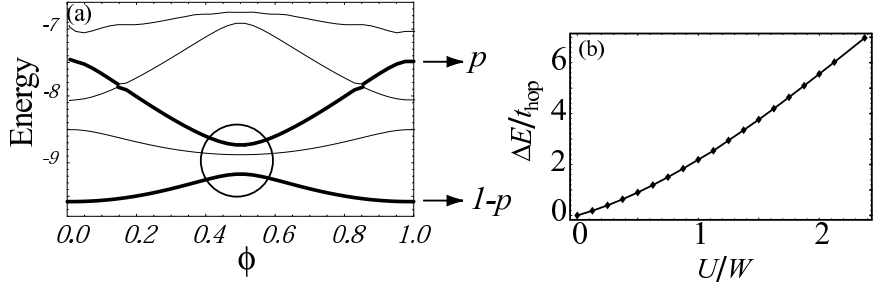}
\caption{
(a) Spectrum of the half-filled Hubbard model $H(\phi)$ for $0\le\phi\le 1$.
Bold lines represent the ground state and the first 
state among the tunneling-allowed excited states, respectively.
(b) $U/W$-dependence of the Mott 
gap $\Delta E$, encircled in (a). 
$W=4t_{\rm hop}$ is the non-interacting band width, 
and system size is $L=10$ and $U/t_{\rm hop}=1.5$.
}
\label{lzHub1}
\end{figure}
In order to calculate the decay rate introduced above, 
we compute the temporal evolution of the 
ground-state survival probability,
\begin{equation}
P_0(t)=
|\bra 0;\phi(s)|\hat{T}e^{-\frac{i}{\hbar}\int_0^tH(\phi(s))ds}|0;0\ket|^2,
\end{equation}
where $|0;\phi\ket$ denotes the ground-state of $H(\phi)$.
The survival probability is related to the 
decay rate of the ground-state by $P_0(t)=e^{-\Gamma t}$.  

The short-time feature in the survival probability is 
expected to be described by the single Landau-Zener transition 
between the ground-state and the 
lowest excited state, displayed by the two
bold lines in the figure\footnote{In Fig.\ref{lzHub1}(a), three
states appear in the circle. However, transition 
from the ground-state to the 
middle state is forbidden by symmetry. 
}, that takes 
place around $\phi=\frac{1}{2}$ (Fig.\ref{lzHub1}(a)). 
If we concentrate on the two levels, 
the time evolution operator at  $t=\Delta t$ 
($\Delta t=\frac{h}{FL}$ is defined as the time 
when $\phi(\Delta t)=1$ is reached) 
is approximated by a $2\times 2$ matrix,
\begin{eqnarray}
U_{\rm{2level}}(t=\Delta t)=
\left(
	\begin{array}{cc}
	\sqrt{1-p}e^{-i\chi}&-\sqrt{p}\\
	\sqrt{p}&\sqrt{1-p}e^{i\chi}
	\end{array}
\right),
\end{eqnarray}
where the tunneling probability $p$ is 
given by the Landau-Zener formula \cite{Landau,Zener,St},
\begin{eqnarray}
&&\hspace{-2cm}
p=\exp\left(-\pi \frac{F_{\rm th}^{\rm LZ}}{F}\right),
\quad F_{\rm th}^{\rm LZ}=\frac{\left[\Delta_c(U)/2\right]^2}{v}.
\label{eq:lzHub}
\end{eqnarray}
Here, $\Delta_c(U)$ is the excitation gap (Fig.\ref{lzHub1}(b)), 
$v=2t_{{\rm hop}}$, and 
$\chi$ the sum of dynamical and Stokes phases. 

\begin{figure}[t]
\centering 
\includegraphics[width=10.cm]{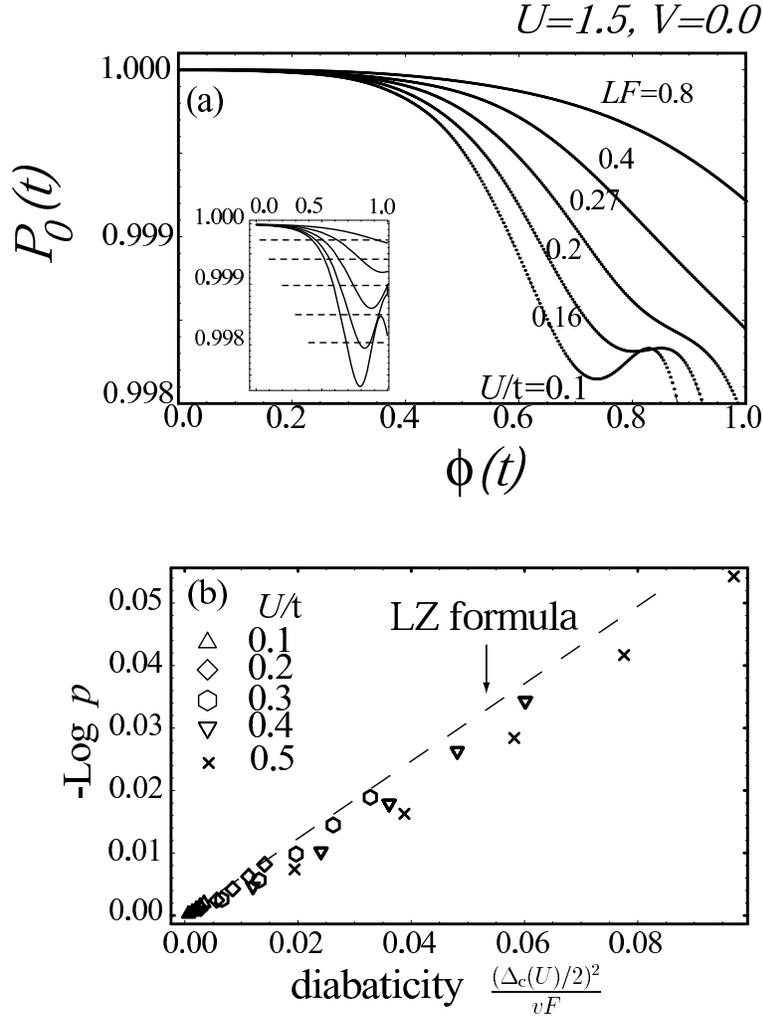}
\caption{
(a) Short-time temporal evolution of the survival probability 
$P_0(t)$ in the half-filled Hubbard model 
($L=10, N_\uparrow=N_\downarrow=5$) for various values of 
$F$ with $U/t=0.1$. The inset shows the 
solutions of the LZS equation 
with its asymptotic values indicated as dashed horizontal lines.
(b)The transition probability $p$, with $P(t=\Delta t)=1-p$, 
plotted against the diabaticity parameter.
The dashed line is the prediction of the Landau-Zener formula.
}
\label{lzHub2}
\end{figure}

In order to verify eq.(\ref{eq:lzHub}),
we can numerically calculate the 
survival probability $P_0(t)$ from $t=0$ to $t=\Delta t$
for various $U$ and $F$ (Fig.\ref{lzHub2}). 
After determining the tunneling probability from 
$p=1-P_0(\Delta t)$, we plot it against the diabaticity
parameter $\frac{(\Delta_c(U)/2)^2}{vF}$. 
The data points(Fig.\ref{lzHub2}(b)) for various values 
of $U$ fall around a common line, which is just 
the prediction of the Landau-Zener formula. 
The agreement is better for smaller values of $U$ where 
we can treat the Umklapp term as a perturbation.

\subsubsection{Long-time behavior --- a time-dependent 
DMRG result}
\begin{figure}[th]
\centering 
\includegraphics[width= 7.2cm]{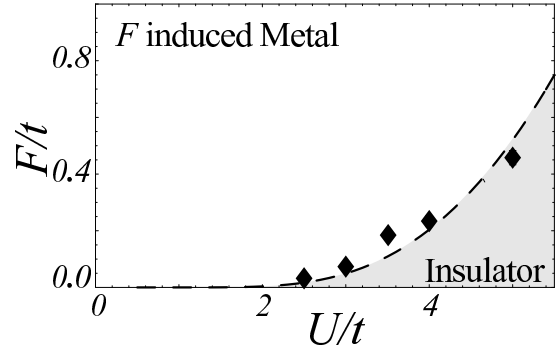}
\caption{The dielectric-breakdown phase diagram 
on the $(U,F)$ plane 
for the one-dimensional Hubbard model. 
The symbols are the threshold $F^{\rm Mott}_{\rm th}(U)$ 
obtained by fitting
the decay rate $\Gamma(F)/L$ to eq.(\ref{fittingfunction})), 
while the dashed line
 is the prediction $F=F_{\rm th}^{\rm LZ}(U)$
of the Landau-Zener formula eq.(\ref{LZMott}).
}
\label{Fth}
\end{figure}

The conclusion obtained in the 
previous section with the exact diagonalization 
is that the short-time 
behavior after the electric field is switched on 
is dominated by the single Landau-Zener transition 
between the ground state and the first excited state.
However, several important questions remain, e.g.,
\begin{description}
\item[Will the first transition remain finite in the infinite-size limit?] 
Indeed, Kohn's criterion (\ref{2KohnCriterion})
asserts that the $\phi$ dependence of the ground-state 
energy of a Mott insulator should vanish for $L\rightarrow \infty$. 
This implies that the 
adiabatic flow (Fig. \ref{fig:adiabaticLevels}(c))
should become flat in this limit, which may seem to indicate 
that the transition will be washed out.  
However, this contradicts with the expression for the 
threshold $F_{\rm Zener}=\frac{[\Delta_c(U)/2]^2}{v}$ (eq.(\ref{eq:lzHub})),
which remains finite in the $L\to \infty$ limit. 
Since this expression is obtained 
in a small system and in the small $U$ limit,
there is a possibility that this breaks down. 
Surprisingly, we shall show that this expression 
survives in large systems even when $U$ is 
not small (Fig. \ref{Fth}).  
\item[The effect of pair annihilation]
After the first transition, we expect the system
to undergo further transitions to higher-energy 
levels.  This process, however, should be couterbalanced by 
another processes, the {\it pair annihilation}
of doublons and holes.  These processes, which do not 
conserve the total momentum in general, are caused by 
the Umklapp scattering. 
Thus the pair creation (= Landau-Zener transitions to 
high-energy states) tends to be offset by pair annihilation, 
which implies that the decay rate of the 
ground state may become smaller compared to the 
single Landau-Zener transition case\footnote{
A similar problem has been studied from a general 
point of view by Wilkinson and Morgan \cite{Wilkinson2000}.}.
\end{description}

These questions have motivated us to study the 
dielectric breakdown in the half-filled Hubbard model
for longer time periods, which is accomplished 
by the time-dependent density matrix renormalization group method. 
A version of the real-time DMRG was first intruduced 
by Cazalilla and Marston with a truncated DMRG 
Hilbert space and a renormalized Hamiltonian \cite{tddmrg}. 
Precision of their method degrades rapidly in the 
long-time limit, since an update of the 
Hilbert space is lacking. Recently, Vidal proposed an improved 
method for simulating time-dependent phenomena in 
one-dimensional lattice systems employing the 
Trotter-Suzuki decomposition \cite{Vidal2003,Vidal2004}. 
White and Feiguin \cite{White2004} as well as other groups \cite{Daley2004}
modified  this idea and combined it with the finite-size 
DMRG algorithm. 

If we denote the DMRG wave function as
\begin{equation}
|\Psi\ket=\sum_{l\alpha_j\alpha_{j+1} r}\psi_{l\alpha_j\alpha_{j+1} r}
|l\ket|\alpha_j\ket|\alpha_{j+1}\ket|r\ket,
\end{equation}
where $|l\ket,\;|r\ket$ is the basis of the truncated
Hilbert space with dimension $m$ and 
$|\alpha_j\ket,\;|\alpha_{j+1}\ket$ are the two sites 
that bridge the left and right blocks in the DMRG procedure. 
By employing the 
Trotter-Suzuki decomposition, 
\begin{equation}
e^{-idt H}\simeq e^{-idtH_1/2}e^{-idtH_2/2}\ldots e^{-idtH_2/2}e^{-idtH_1/2},
\end{equation}
we can apply the time-evolution operator $e^{-idtH_j/2}$ 
to the $j$-th wave function as 
\begin{equation}
\left(e^{-idtH_j/2}\psi\right)_{l\alpha_j\alpha_{j+1} r}=
\sum_{\alpha_j'\alpha_{j+1} '}
\left(e^{-idtH_j/2}\right)_{\alpha_j\alpha_{j+1};\alpha_j'\alpha_{j+1}'}
\psi_{l\alpha_j'\alpha_{j+1}' r} .
\end{equation}
After applying $e^{-idtH_j/2}$, we 
diagonalize the density matrix and move to the next link 
just as in the usual finite-size algorithm. 
One cycle of this procedure results in an evolution of 
time by $dt$, and we can repeat it as many times as we wish. 
Compared with the version by Cazalilla-Marston \cite{tddmrg}, 
this algorism has higher precision 
and we can simulate non-equilibrium 
excited states efficiently\cite{Daley2004}, 
although one drawback of the t-dependent DMRG is that 
we can only treat systems with open boundary conditions.

Here we study transient behaviors of the 
one-dimensional Hubbard model with open boundary condition. 
We use the time-independent gauge, for which 
the Hamiltonian is 
\begin{equation}
H(F)=-t_{\rm hop}\sum_{j,\sigma}\left(c_{j+1\sigma}^\dagger 
c_{j\sigma}+\mbox{h.c.}\right)+U\sum_j n_{j\up}n_{j\dw}+F\hat{X},
\end{equation}
where $\hat{X}=\sum_jjn_j$ is the position operator 
representing the tilted potential. 
As in the previous section, we start the time evolution 
from the $F=0$ 
ground-state $|0\ket$ obtained by the usual 
finite-size DMRG.  The wave function in this gauge is simply
\begin{equation}
|\Psi(t)\ket=e^{-itH(F)}|0\ket,
\end{equation}
which is obtained with the t-dependent DMRG.

\paragraph{Evolution of the charge density}
\begin{figure}[ht]
\centering 
\includegraphics[width=12cm]{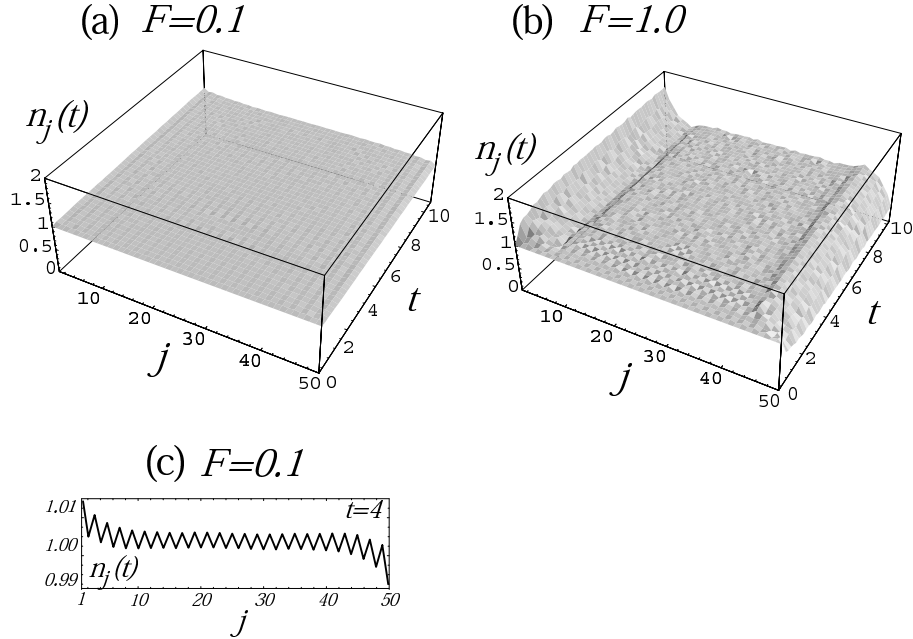}
\caption{
Temporal evolution of the charge density $n_j(t)$ in the 
half-filled Hubbard model with $m=150, L=50, dt=0.02$ for 
$F=0.1$(a) and $F=1.0$(b).  (c) depicts a cross section 
of (a) for $t=4$.
}
\label{densityU4}
\end{figure}

We first discuss the temporal evolution of the 
charge density, 
$n_j(t)=\bra \Psi(t)|n_j|\Psi(t)\ket$, 
after the electric field is switched on at $t=0$. 
At half-filling the initial distribution is $n_j(t)=1$. 
After the application of the electric field, 
a charge density wave (CDW) pattern is formed 
when the electric field is not too strong (Fig. \ref{densityU4}(a)).
This state is stationary and the density profile do 
not change any further. 
The pattern is formed because the boundary condition 
breaks the translational symmetry, where the amplitude of the pattern 
corresponds to the polarization $\Delta P(F)$ induced
by the field.  
When the electric field becomes stronger, 
charge transfers start to occur, with charge 
accumulation and charge depletion being formed 
around the edges in an open-boundary chain (Fig.\ref{densityU4}). 
This is a sign that the ground state collapses 
due to quantum tunneling.

\paragraph{The decay rate of the ground state}
The groundstate-to-groundstate transition amplitude is, in the 
time-independent gauge, 
\begin{equation}
\Xi(t)=\bra 0|e^{-\frac{i}{\hbar} \tau\left(H+F\hat{X}\right)}|0\ket e^{\frac{i}{\hbar} t E_0},
\label{eq:transitionamplitude2}
\end{equation}
where we denote the ground state of $H$ as $|0\ket$ 
and its energy as $E_0$. 
\begin{figure}[t]
\centering 
\includegraphics[width= 9.cm]{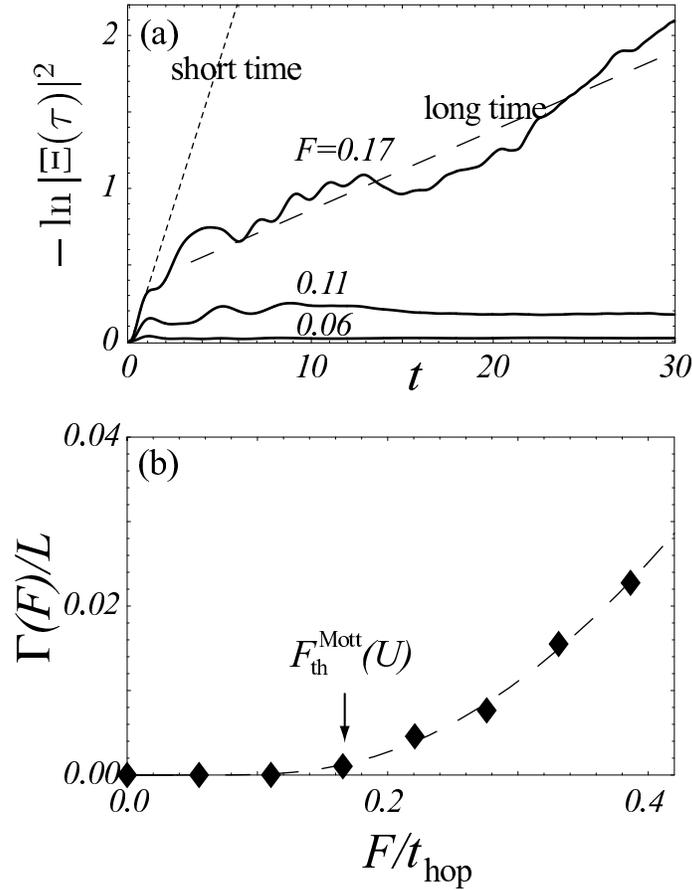}
\caption{
(a) Temporal evolution of the ground-state survival 
probability $|\Xi(t)|^2$ 
after the electric field $F$ is switched on at $t=0$ 
in the 1D half-filled Hubbard model with $U/t_{\rm hop}=3.5$, 
obtained with the time-dependent DMRG for $L=50$ 
with the size of the DMRG Hilbert space $m=150$
and the time step $dt=0.02$.
The dashed line represents $-\ln|\Xi(t)|^2=\Gamma(F)t+c$ 
for $F/t_{\rm hop}=0.17$, while the dotted line 
delineates the initial slope (the short-time behavior).  
(b) The decay rate versus $F$
in the half-filled Hubbard model.
Dashed curve is a fit to eq.(\ref{fittingfunction}),
where $F_{\rm th}^{\rm Mott}(U)$ is the 
threshold.
}
\label{ImSU35}
\end{figure}
Figure \ref{ImSU35}(a) shows the temporal evolution of the 
ground-state survival probability $|\Xi(t)|^2$ 
for a system with $U/t_{\rm hop}=3.5$.  
As time evolves, 
the slope of $-\ln |\Xi(t)|^2$ ($\propto$ the decay rate) 
decreases after an initial stage, 
which implies a suppression of the tunneling from the 
short-time behavior. 
This should indicate that charge excitations 
are initially 
produced due to the Landau-Zener tunneling
from the ground state to the first excited states, 
but that scattering among the excited states 
become important as the population of the excitations 
grows. In other words, pair annihilation of carriers
becomes important and acts to suppress the tunneling rate. 
We have determined $\Gamma(F)$ 
from the long-time behavior with a fitting
$-\ln |\Xi(t)|^2=\Gamma(F)t+{\rm const}$. 

The decay rate per length $\Gamma(F)/L$ 
is plotted in  Fig.\ref{ImSU35}(b), 
where we have varied the system size ($L=30, 50$) 
to check the convergence.
$\Gamma(F)/L$ is seen to remain vanishingly small until 
the field strength exceeds a threshold. 
To characterize the threshold $F_{\rm th}(U)$ 
for the breakdown we can evoke the form obtained above 
for the one-body system.  
The formula (eq.(\ref{BandSchwingerFormula}) for $d=1$ 
with the error function ignored and 
the factor of $2$ recovered for the spin degeneracy), 
\begin{equation}
\Gamma(F)/L=-\frac{2F}{h}a(U)\ln
\left[1-\exp\left(-\pi \frac{F_{\rm th}^{\rm Mott}(U)}{F}\right)\right],
\label{fittingfunction}
\end{equation}
is originally derived for one-body problem, 
and an obvious interest here is whether the formula 
can be applicable if we replace the one-body 
$F_{\rm th}^{\rm band}$ with the many-body $F_{\rm th}^{\rm Mott}(U)$.
In the above we have added a factor $a(U)$, a parameter 
representing the suppression of the quantum tunneling. 
The dashed line in Fig.\ref{ImSU35}(b) is the 
fitting to the formula for $U/t_{\rm hop}=3.5$, 
where we can see that the fitting, including the essentially 
singular form in $F$, is surprisingly good, 
given a small number of fitting parameters. 
The value of $a(U)$ turns out to be close to but smaller than unity 
(taking between $0.77$ to $0.55$ as $U/t$ is increased from $2.5$ to $5.0$).

If we perform this for various values of $U$ we can construct 
a ``nonequilibrium (dielectric-breakdown) phase diagram", 
as displayed in 
Fig.\ref{Fth}, which plots the $U$ dependence of $F_{\rm th}^{\rm Mott}$.
The dashed line 
is the prediction of the Landau-Zener formula \cite{Oka2003},
\begin{equation}
F_{\rm th}^{\rm LZ}(U)=\frac{[\Delta_{\rm c}(U)/2]^2}{v}.
\label{LZMott}
\end{equation}
For the size of the Mott (charge) gap we use the 
Bethe-ansatz result,\cite{Lieb:1968AM}
\begin{equation}
\Delta_{\rm c}(U)=\frac{8t_{\rm hop}}{U}\int_1^\infty\frac{\sqrt{y^2-1}}
{\sinh(2\pi yt_{\rm hop}/U)}dy, 
\end{equation}
with $v/t_{\rm hop}=2$.  As can be seen, the DMRG result 
and the Landau-Zener result agrees surprisingly well.

\subsection{Long-time behavior and a mapping to a quantum random walk}
\begin{figure}[htb]
\centering 
\includegraphics[width=11cm]{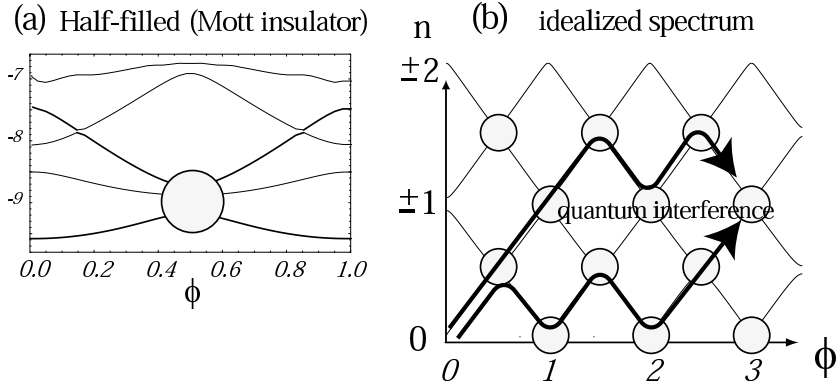}
\caption{
(a) The spectrum of the half-filled Hubbard model.
The circle corresponds to Landau-Zener transition
between the two energy levels which can
be expressed by a $2\times 2$ unitary matrix
(eq.(\ref{eq:lzunitarywalk})).
(b) Idealized energy levels where 
level anti-crossings are 
expressed by circles. 
Quantum interference takes place when 
contributions from different paths are considered.
}
\label{LZ2QW}
\end{figure}

Since many levels should be involved in the above 
pair creation/annihilation processes, next thing we want to 
have is a statistical mechanical setup for 
the time evolution of the Mott insulator.  
The problem at hand is a closed quantum system in external driving forces
(e.g., electric fields), which are represented by a 
time varying parameter $\phi(t)$ of the Hamiltonian. 
We want to discuss the asymptotic solution of the 
time-dependent Schr\"odinger equation
\begin{equation}
i\hbar\frac{d}{dt}|\Psi(t)\ket=H(\phi(t))|\Psi(t)\ket.
\end{equation}
We introduce $|n;\phi\ket$ as the set of eigenstates of the 
time-dependent Hamiltonian $H(\phi)$, and denote the 
energy eigenvalue as $E_n(\phi)$, i.e., 
$H(\phi)|n;\phi\ket=E_n(\phi)|n;\phi\ket$ (Fig.\ref{LZ2QW}). 
Since $|n;\phi\ket$ forms a complete orthonormal basis, 
the wave function $|\Psi(t)\ket$ can be expanded as
\begin{equation}
|\Psi(t)\ket=\sum_n\psi(n,t)
e^{-\frac{i}{\hbar}\int_0^tE_n(s)ds}
|n;\phi(t)\ket
\label{5WaveVector}
\end{equation}
with coefficients
$\psi(n,t)
=\bra n;\phi(t)|\hat{T}e^{-i\int_0^tH(\phi(s))ds}|0;0\ket/e^{-\frac{i}{\hbar}\int_0^tE_n(s)ds}$. 
Note that we have removed the contribution from the 
dynamical phase $\int_0^tE_n(s)ds/\hbar$ 
in the definition of  $\psi(n,t)$. Although the evolution 
depends on the  detail of the system ($H(\phi)$), 
we can
deduce some {\it universal} features that depend only on the 
feature of the energy levels, i.e., distribution of level repulsion 
in the spectrum. 

Each energy level is subject to the Landau-Zener tunneling 
to neighboring levels in a time period $\Delta t/2$, 
and is most 
conveniently expressed in terms of the transfer matrix
representation\cite{Oka2004a,Nakamura1987}. 
To this end, we denote the pairs as
\begin{equation}
\Psi(n,\tau)= \left(
	\begin{array}{c}
	\psi_L(n,\tau)\\
	\psi_R(n,\tau))	
	\end{array}
\right),
\end{equation}
and the time evolution ``rule" can be expressed as
\begin{equation}
\Psi(n,\tau+1)=P_{n+1}\Psi(n+1,\tau)+Q_{n-1}\Psi(n-1,\tau),
\label{BQW1}
\end{equation}
where $P_n$ ($Q_n$) is  
the upper (lower) half 
of a $2\times 2$ unitary matrix,
\begin{equation}
U_n=\left(
	\begin{array}{cc}
	a_n&b_n\\
	c_n&d_n
	\end{array}
\right),\quad 
P_n=\left(
	\begin{array}{cc}
	a_n&b_n\\
	0&0
	\end{array}
\right),\quad 
Q_n=\left(
	\begin{array}{cc}
	0&0\\
	c_n&d_n
	\end{array}
\right).
\end{equation}

\begin{figure}[tb]
\centering 
\includegraphics[width=3.5cm]{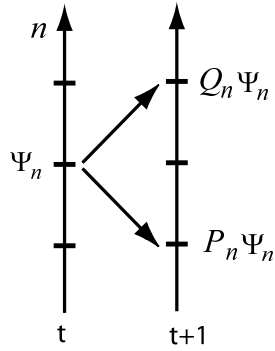}
\caption{
Application of $Q_n$ and $P_n$ in the quantum random walk.
}
\label{5PandQ}
\end{figure}
The diagonal elements of 
$U_n$ represent the  Landau-Zener transition 
from  the $n$-th level to $(n-1)$ or $(n+1)$-th level,
where the explicit form is  
\begin{equation}
U_n=
\left(
	\begin{array}{cc}
	\sqrt{p_n}e^{i\beta_n}&\sqrt{1-p_n}e^{i\gamma_n}\\
	-\sqrt{1-p_n}e^{-i\gamma_n}&\sqrt{p_n}e^{-i\beta_n}
	\end{array}
\right).
\label{eq:lzunitarywalk}
\end{equation}
Here the Landau-Zener tunneling probability $p_n$ 
depends on the ratio of the Zener threshold field $F_{\rm Zener}^n$ and 
the electric field $F$ as
\begin{equation}
p_n=\exp\left(-\pi\frac{F_{\rm Zener}^n}{F}\right),
\end{equation}
where $F_{\rm Zener}^n$ generically depends on $n$. 

If we regard $\Psi(n,\tau)=\left(
	\begin{array}{c}
	\psi_L(n,\tau)\\
	\psi_R(n,\tau)
	\end{array}
\right)$
as a ``qubit" on ``site" $n$, eq.(\ref{BQW1})
defines an evolution of a {\it one-dimensional quantum walk}
with a  reflecting boundary 
at $n=0$ corresponding to the ground-state (Fig.\ref{LZ2QW}(c)).
A quantum walk is a quantum counterpart of the classical 
random walk.  
Models with essentially equivalent ideas have appeared 
in various fields: 
to name a few, quantum transport and dissipation \cite{Blatter1988,Oka2003}, 
quantum Hall effect \cite{Chalker1988},
optics \cite{Bouwmeeste2000,Wojcik2004}
and recently in quantum information
\cite{Nayak2000,Konno2003,Konno2002b,KonnoReviewPQRS,Mackay2002,KonnoNamikiSoshi,
Konno2002a,Bach2002,Yamasaki2003,Tregenna2003, Kempe2003,
Tregenna2003,Ambainis2003}.
In the field of quantum information (see e.g. \cite{NielsenBook}),
introduced by Aharonov, Ambainis, Kempe and Vazirani
in 2001 \cite{Aharonov2001}, the quantum walk is 
arousing interest in hope of revealing new features in the 
quantum algorithms (for reviews see \cite{Kempe2003,Tregenna2003,Ambainis2003}). 
Researches stem into many directions, e.g., the effect of absorbing boundary 
conditions \cite{Bach2002,Yamasaki2003,Konno2003}, 
higher-dimensional systems \cite{Mackay2002,Tregenna2003,Inui2004twodim}, 
localization in systems with internal degrees of freedom \cite{Inui2004multi},
and many powerful analytical techniques are being developed.

\begin{figure}[h]
\centering 
\includegraphics[width=11.5cm]{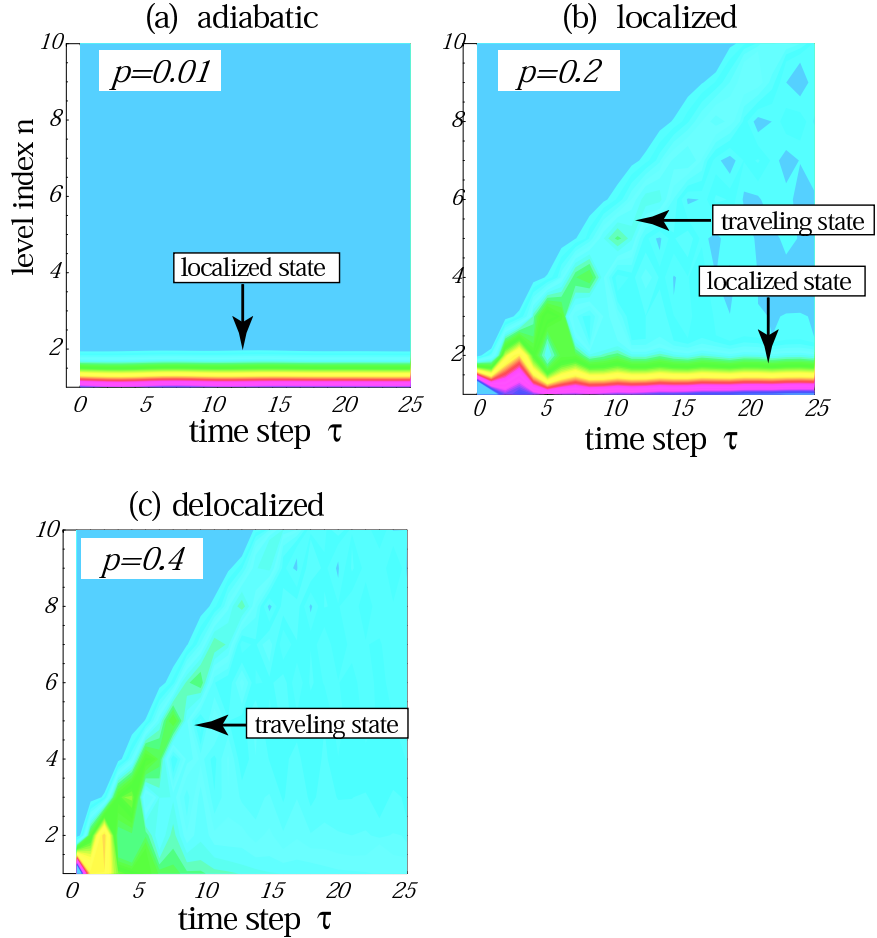}
\caption{
The time evolution of the  distributions of wave function
amplitude $\rho(n,t)=|\psi(n,t)|^2$ in energy space.
The vertical axis $n$ is the index of the energy levels.
(a)For small tunneling, 
the distribution is localized at the ground state.
(b) For intermediate tunneling,
a localized state remains, while the amplitude starts to 
bifircate into excited states
of the wave function is excited.
(c) When the tunneling is larger 
than the threshold, the localized state 
disappears.
}
\label{1ThreeAsymptotics}
\end{figure}
An important feature of the quantum walk, as opposed to the 
classical walk, is that different transition paths 
interfere with each other quantum mechanically.  
We in fact find that the quantum interference 
leads to a {\it dynamical localization}, 
an analog of Anderson's localization taking place in the 
energy space rather than in the position space. 
In our previous work \cite{Oka2004a} we have employed the 
PRQS method, a technique to treat quantum walks, 
to perform the path integral, 
and obtained the exact asymptotic distributions of the wave 
function for a simplified model. 
The resultant states can be categorized in 
three types depending on the strength of the electric field, 
as  schematically plotted in Fig. \ref{1ThreeAsymptotics}.
(A) is an adiabatic evolution that takes 
place in weak driving forces (electric fields).
The dominant part of the distribution $\rho(n,t)$ against energy 
is a delta function localized around the ground-state.
When the driving force  
become stronger, quantum tunneling broadens the 
delta function, as plotted in (B). 
The shape of the peak is maintained 
by a balance between tunneling and dynamical localization. 
(C) is the case where the driving force overwhelms the 
effect of dynamical localization, 
and the system is driven rapidly into the excited states. 
This, in our view, corresponds to the dielectric breakdown.

\newpage

\subsection{Experimental implications}
\label{3Experimentalresults}
Now we discuss experimental implication of 
the many-body Landau-Zener transition mechanism.
In fact, there are several mechanisms 
which may lead to breakdown of insulators. 
For example, Fr\"{o}hlich's electric avalanche mechanism
may take place, in which a small number of 
excited electrons act as a seed and become accelerated by
the electric field until they cooperatively 
destroy the insulator.  
We can distinguish Landau-Zener transition 
from the avalanche mechanism through the 
temperature dependence and from interface effects
by changing the size of the sample.
Another important effect is 
the band bending near an interface of the Mott insulator and  
electrodes. 
For a thin sample, this may lead to 
injection of carriers, and results in the 
interface Mott transition\cite{OkaNagaosa}.

\begin{figure}[htb]
\centering 
\includegraphics[width= 12.cm]{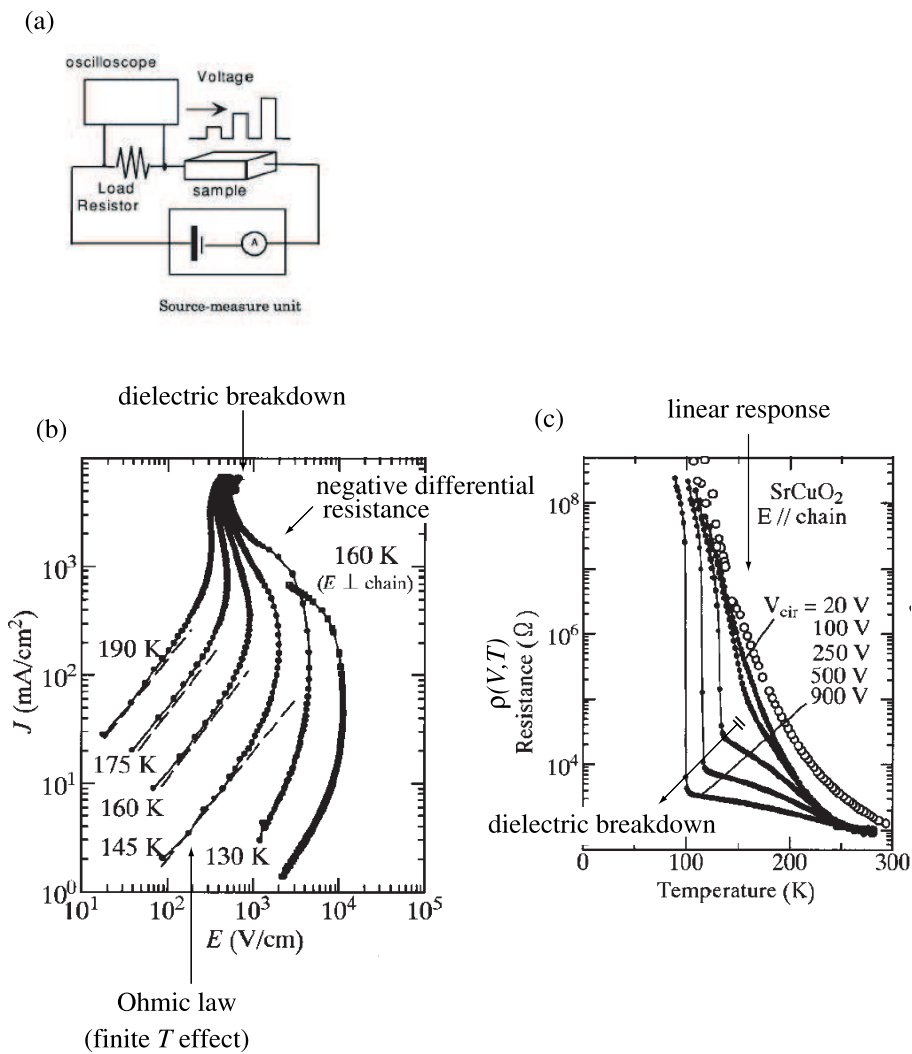}
\caption{
(a) A schematic measurement circuit.
(b) $J-E$ characteristics of a SrCuO$_2$ sample.
(c) Temperature dependence of the resistance for 
various applied voltages. 
After Taguchi {\it et al.} \cite{tag}. 
}
\label{4IVExp}
\end{figure}

Dielectric breakdown of one-dimensional Mott insulators
was experimentally studied by Taguchi {\it et al.}, 
who obtained the $J-E$ characteristics (Fig.\ref{4IVExp}) of 
Sr$_2$CuO$_3$ and  SrCuO$_2$ samples\cite{tag}, 
which are both quasi-1D, strongly correlated electron systems.  
Experiments were done by placing small single crystals 
in circuits as shown in Fig.\ref{4IVExp}(a), 
and the voltage drop $V$ was measured while the
current density $J$ was fixed.
Depending on the strength of the electric field $F$, 
transport properties change drastically, 
as summarized in the following.

\begin{figure}[htb]
\centering 
\includegraphics[width= 11.5cm]{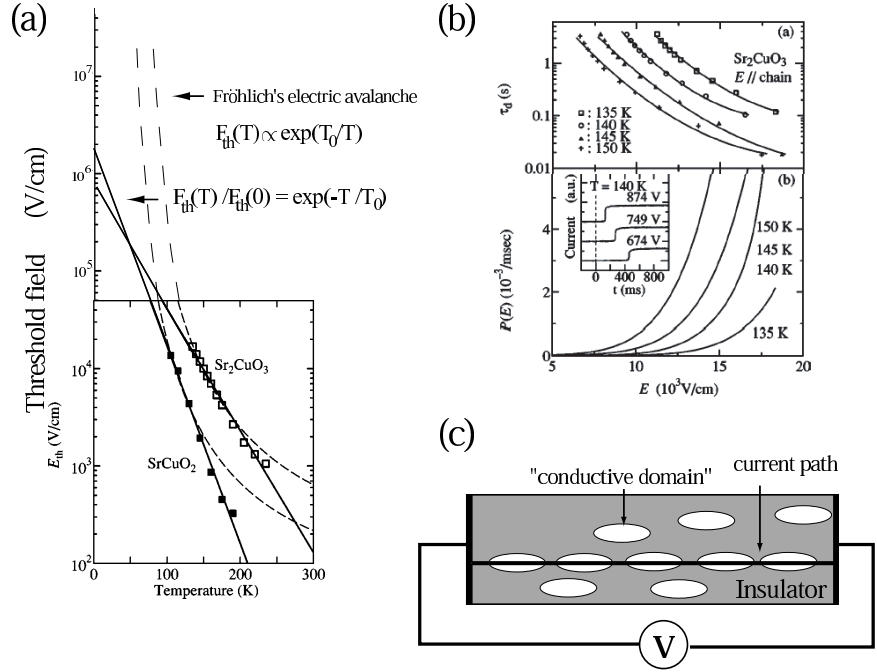}
\caption{
An experimental result for the
temperature dependence of the 
threshold electric field $F_{\rm th}(T)$ 
for the dielectric breakdown (adopted from Taguchi {\it et al.} \cite{tag}). 
Dashed lines correspond to a fitting to $\exp(T_0/T)$
predicted by Fr\"ohlich's electron avalanche.  
Solid lines correspond to a fitting to $\exp(-T/T_0)$. 
(b) The electric-field dependence of the delay time $\tau_{\rm d}$
for Sr$_2$CuO$_3$. 
(c) The production rate of ``conductive domains".
The inset exemplifies the temporal evolution of the 
current at various applied voltages.
Adopted from Taguchi {\it et al.} \cite{tag}.
}
\label{4FTemperature}
\end{figure}

In weak electric fields, the $J-E$ 
characteristics shows an Ohmic behavior
at finite temperatures. 
When the electric field exceeds
a threshold value, the current 
shows a dramatic increase.
Such drastic changes cannot be explained by 
perturbation in $F$, and we must 
consider non-perturbative effects, i.e., a behavior 
essentially singular in $F$ like
$J\sim\mbox{function of} \exp\left(-F_{\rm th}/F\right),$ 
which is a typical tunneling effect
with threshold $F_{\rm th}$.
The temperature dependence (Fig.\ref{4FTemperature}) of the 
threshold can be fit well by 
$F_{\rm th}(T)/F_{\rm th}(0)\sim \exp\left(-T/T_0\right)$.
This excludes the avalanche mechanism, for which an 
activation type temperature dependence
($F_{\rm th}^{\rm avalanche}(T)/F_{\rm th}(0)\sim \exp\left(T_0/T\right)$)
is expected.  

One indication that the breakdown is indeed quantum in nature 
is that the threshold extrapolates to 
a finite value for $T \rightarrow 0$.  
From the extrapolation (Fig.\ref{4FTemperature}(a)), 
we obtain a threshold,
\begin{equation}
F_{\rm th}^{\rm exp}\sim 10^6-10^7\;(\mbox{eV/cm}),
\end{equation}
for Sr$_2$CuO$_3$ and  SrCuO$_2$.
The Landau-Zener result (intended for $T=0$) of the 
threshold (eq.(\ref{eq:lzHub})) is
\begin{equation}
F_{\rm th}^{\rm LZ}(U)=\frac{[\Delta_{\rm c}(U)/2]^2}v
\sim (1\mbox{eV})^2/(10^{-7}\mbox{eV/cm}) \sim 
10^6\;(\mbox{eV/cm})
\end{equation}
is comparable with the experimental result. 

Interestingly, the decay rate $\Gamma(U)$ we have introduced 
theoretically can be 
measured experimentally\cite{tag}.
This is done by 
studying the  transient behavior of the current
after the electric field is switched on at $t=0$.
At first the current density is zero, 
and then becomes non-zero after 
a certain delay time 
$t=\tau (F)$(Fig.\ref{4FTemperature}(b), lower inset).
The authors in \cite{tag} 
have introduced a phenomenological percolation model to 
relate the delay time with the production rate $P(F)$ of 
the conductive domains (see \cite{tag}). 
In this model, 
conductive domains are envisaged to grow in the sample, 
and the current density is assumed to become finite 
when the left and right electrodes are 
connected by these domains (Fig.\ref{4FTemperature}(c)).
This leads to a relation,
\begin{equation}
P(F)=-\left(F\frac{d\tau}{dF}\right)^{-1}.
\end{equation}
The experimental result for the production rate in 
Fig.\ref{4FTemperature}(b) is obtained in this way.

The nature and the microscopic origin of the 
``conductive domain" are not clear, but if we interpret 
them to be domains with a high density of charge excitations
produced by the Landau-Zener transition, 
the vacuum decay rate per volume
$\Gamma(F)/L^d$ characterizes 
quantum tunneling from the ground-state
to excited states.  
With the identification we expect that the decay 
rate and the production rate are identical, i.e.,
\begin{equation}
P(F)\sim \Gamma(F).
\end{equation}
This identification is encouraged by the
field dependence of $P(F)$ (lower panel of 
Fig.\ref{4FTemperature}(b)), which is close to the 
expected form ($\Gamma/L\sim
-\frac{2F}{h}a(U)
\ln\left[1-\exp\left(
-\pi\frac{F_{\rm th}^{\rm Mott}}{F}\right)
\right]$) of the decay rate.

In this experiment a scaling study --- 
a systematic change of the size of the sample --- 
was also performed to confirm that the nonlinear 
effect occurs in the bulk. 
From these observations, we conclude that 
the experiment by Taguchi {\it et al.}\cite{tag} can be 
explained by the many-body Landau-Zener tunneling mechanism.
However, to be more confident, we need to know the temperature 
dependence of the threshold theoretically, which is still a challenging 
task in the present many-body system.

\subsection{Conclusion}
In  this article, we have  
explained how dielectric breakdown of Mott insulators
can be explained from the nonequilibrium behaviors of 
charge carriers, especially from their 
creation and annihilation processes.  
Both processes are the result of many-body Landau-Zener
nonadiabatic tunneling transition between  many-body energy levels, 
where charge creation processes are counterbalanced 
by annihilation processes. 
From numerical result we have obtained a nonequilibrium 
(dielectric-breakdown) phase diagram.  
If the coherence of the dynamics is preserved 
at sufficiently low temperatures, a quantum interference, 
as modeled by a quantum walk in energy space, 
may lead to dynamical localization, which 
saturates the creation process and 
leads to a non-equilibrium stable state. 
The decay rate $\Gamma(F)$ that we have discussed 
is a measurable quantity: it is 
the  production rate observed by Taguchi {\it et al.} \cite{tag} 
in copper oxides. The experimental result is consistent with 
our prediction $\Gamma(F) \sim \frac{F}{2\pi}\ln (1-p)$
when an extrapolation to zero temperature is made.
It is an interesting future topic to 
understand the properties of the 
non-equilibrium stable state in more detail.  

An important open question is how the energy dissipation 
processes take place in nonequilibrium situations.  
Here we have stressed that the many-body processes 
act effectively as a source of dissipation 
through scattering, but 
an explicit incorporation of 
heat-bath effects, electrode effects, etc, 
is left to a future problem.  

{\it Acknowledgements:}
We wish to thank Rytaro Arita and Norio Konno for the collaboration 
in the workes described here and for illuminating discussions.
We also indebted to Yshai Avishai and Paul Wiegmann for 
illuminating discussions.
TO acknowledges Masaaki Nakamura, Kazuma Nakamura, 
Shuichi Murakami, and Naoto Nagaosa 
for helpful comments.


\end{document}